\documentclass[11pt]{article}
\usepackage{arxiv}

\usepackage{fullpage}
\usepackage{framed}
\usepackage{mdframed}

\usepackage{graphicx}
\usepackage{caption}
\usepackage{subcaption}
\usepackage{wrapfig}
\usepackage{svg}

\usepackage{amsmath}
\usepackage{amssymb}
\usepackage{amsthm}
\usepackage{mathrsfs} % Fancy scripted font
\usepackage{bm}  % Bold math
\usepackage{centernot}  % \centernot\implies looks better than \not\implies

\usepackage[linesnumbered, ruled, vlined]{algorithm2e}
\usepackage{soul}  % \hl highlighting
\usepackage{color}
\usepackage{mathtools}  % For my \ceil function

\usepackage[backend=biber,sorting=nyt,giveninits=true,maxnames=10000]{biblatex}
\usepackage{enumitem}
\usepackage{etoolbox}
\usepackage{hyperref}
\usepackage{todonotes}
\bibliography{refs}
\usepackage{booktabs}
\usepackage{thmtools}
\usepackage{environ}
\theoremstyle{plain}  % Style definition removes italics

\usepackage[conf={restate}]{proof-at-the-end}

\theoremstyle{definition}
\newtheorem{remark}{Remark}[section]
\newtheorem{definition}{Definition}[section]

\newtheorem{assumption}{Assumption}[section]

\def\defeq{\overset{\Delta}{=}}  % Equal with triangle
\def\d{\mathsf{d}}  % Differential operator
\def\L{\mathcal{L}}  % Lagrangian

\def\E{\mathbb{E}}  % Expectation
\DeclareMathOperator\Var{\text{Var}}  % Variance
\def\P{\mathbb{P}}  % Probability Measure
\def\R{\mathbb{R}}  % Set of real numbers
\def\T{\mathsf{T}}  % Transpose notation
\def\ind{\bm{1}}  % Ones vector or indicator
\def\<{\langle}  % < Inner product
\def\>{\rangle}  % > Inner product

\def\ie{\emph{i.e.,}}
\def\eg{\emph{e.g.,}}
\def\cf{\emph{c.f.,}}

\def\etc{\emph{etc.}}
\def\iid{\emph{i.i.d.}}

\def\cdf{\emph{c.d.f.}}

\def\B{\mathcal{B}}  % Contract set B_j
\def\A{\mathcal{A}}  % Targetting set A_i

\graphicspath{{./figures/}}

\title{Convexity and Duality in Optimum Real-time Bidding and Related Problems}

\author{R. J. Kinnear\\
  \small\href{mailto:ryan@kinnear.ca}{ryan@kinnear.ca} \and
  R. R. Mazumdar\\
  \small\href{mailto:mazum@uwaterloo.ca}{mazum@uwaterloo.ca} \and
  P. Marbach\\
  \small\href{mailto:marbach@cs.utoronto.edu}{marbach@cs.utoronto.edu}}

\begin{document}
\maketitle
\begin{abstract}
  We study problems arising in real-time auction markets, common in e-commerce and computational advertising, where bidders face the problem of calculating optimal bids.  We focus upon a contract management problem where a demand aggregator is subject to multiple contractual obligations requiring them to acquire items of heterogeneous types at a specified rate, which they will seek to fulfill at minimum cost.  Our main results show that, through a transformation of variables, this problem can be formulated as a convex optimization problem, for both first and second price auctions.  Convexity results in efficient algorithms for solving instances of this problem, and the resulting duality theory admits rich structure and interpretations.  Additionally, we show that the transformation of variables used to formulate this problem as a convex program can also be used to guarantee the convexity of optimal bidding problems studied by other authors (who did not leverage convexity).  Finally, we show how the expected cost of bidding in second price auctions is formally identical to certain transaction costs when submitting market orders in limit order book markets.  This fact is used to analyze a Markowitz portfolio problem which accounts for these transaction costs, establishing an interesting connection between finance and optimal bidding.

\end{abstract}

\paragraph{Keywords}
Real-time Bidding; Computational Advertising; Auction Theory; Second Price Auction; Convex Optimization

\paragraph{Acknowledgement}
We acknowledge the support of the Natural Sciences and Engineering
Research Council of Canada (NSERC), [funding reference number
518418-2018].  Cette recherche a été financée par le Conseil de
recherches en sciences naturelles et en génie du Canada (CRSNG),
[numéro de référence 518418-2018].

\clearpage
\tableofcontents
\clearpage

\section{Introduction}
\label{sec:introduction}
Auctions are effective mechanisms for facilitating efficient transactions between buyers and sellers of certain types goods and services.  Auction theory is an important sub-field of applied economics and its importance was recognized in the award of the Sveriges Riksbank Prize in Economic Sciences in Memory of Alfred Nobel to W. Vickrey in 1996 and its place firmly cemented in economic theory with the award of the prize to P. Milgrom and R. Wilson in 2020.  Auctions are in essence allocation mechanisms and have been applied to a number of important problems, namely: the allocation of radio spectrum~\cite{milgrom1998game}, in the organization of electricity markets~\cite{wilson2002architecture, wang2015review}, and are, of course, common mechanisms for the sale of art and antiques, \etc\  Auction markets have also come to play an important role in e-commerce and computational advertising with \textit{sponsored search} and \textit{real-time bidding} (RTB) forming major revenue sources for some of the largest businesses in the world~\cite{edelman2007internet, wang2017display, iab2020iab, choi2020online}.  This paper focuses on real-time bidding, but also explores some connections to finance in the cost of placing market orders and in the problem of optimal portfolio construction.

Briefly, real-time bidding is a market for matching buyers and sellers of a random sequence of heterogeneous items through rapid sealed-bid auctions.  In the context of computational advertising, the items being auctioned are \textit{impressions}.  Specifically, when someone (a \textit{user}) visits a website, information about the visitor is sent to an auction house and advertisers bid for the right to have their content displayed to the user by the website; this display is called an impression.  If the user decides to click on the ad then it is called a \textit{click}, and if they also choose to take further action (\eg~make a purchase) then it is called a \textit{conversion}.  The seller of the item may set some \textit{reserve price} (\ie~the minimum bid that can win the item), but otherwise the winner is always the highest bidder.  However, what they actually pay depends upon the auction mechanism: in a \textit{first price auction} the winner pays what they bid, and in a \textit{second price auction} (also called a \textit{Vickrey auction}~\cite{vickrey1961counterspeculation}), they pay the highest competing bid, which we will call the \textit{price}.  These are the most common auction mechanisms in practice, and we focus on them exclusively.

The problem studied in this paper is the problem of optimal bidding faced by firms that operate as intermediary demand aggregators called \textit{Demand Side Platforms} (DSPs).  Specifically, we study a \textit{contract management} problem where the intermediary has sold contracts which obligate them to acquire some specified quantities of items through the RTB market in exchange for a fee.  Since these are contractual obligations, the DSP will not be faced with a budget constraint, and will thus seek to fulfill the contracts at minimum cost.

There is a natural economic niche occupied by the DSPs, analyzed formally by~\cite{balseiro2017optimal}, see also the discussion of~\cite{zhang2015statistical}.  DSPs offer the technical infrastructure and talent necessary to effectively participate in the online advertising space as small firms are simply unable to afford this expense and most large businesses can be expected to benefit from outsourcing this capacity due to comparative advantages.  The intermediary may also serve a risk hedging function.  Indeed, the contract, once accepted, \textit{must} be fulfilled, even if it is no longer profitable for the DSP to do so, \ie~the DSP bears the risk of adverse changes in the market.  Finally, as is explained by~\cite{balseiro2017optimal}, since the intermediary is not budget constrained, they have at their disposal a wider array of bidding strategies than does a budget-constrained individual advertiser, and this leads to cost reductions and profitability even if DSPs are not technologically more sophisticated.

The problem formulation studied in this paper was first presented in ~\cite{marbach_bidding_2020}.  These ideas were developed further in~\cite{kinnearbidding2_2020} which this current paper builds upon.  There are two key distinguishing factors between the models of these papers, and earlier papers on optimum bidding (see \eg~\cite{ghosh2009adaptive, chen2011real, lang2012handling, zhang2014optimal, karlsson2014adaptive, cai2017real, wu2018budget} for a representative selection): the lack of a budget constraint, and that the constraints require an \textit{allocation} decision (\ie~if an item is won, towards which contract should it be allocated?).  Budget constrained problems are in some sense dual to our model: the goal in the former is to maximize the value of items obtained subject to the budget constraint, and the goal of the latter is to fulfill contract constraints at minimum cost.  The decision about which contract towards which to allocate items is analogous to a network transportation problem --- a similar analogy can be recognized in~\cite{chen2005efficient} which explicitly involves a transportation network, and in~\cite{zhang2015statistical} where similar allocation decisions arise in a different context; we remark upon this paper further in Section~\ref{sec:related_problems}.

Randomness in this model arises from two sources: the arrival of requests, and the competing bidders who wish to display their ads. A reasonable assumption about the bids or prices is that they are \iid\ from some unknown distribution. This can be understood as a consequence of a mean-field competitive equilibrium.  Formal analysis is presented in ~(\cite{iyer2014mean, balseiro2015repeated, balseiro2019learning})  where, in stylized bidding markets, there exist competitive equilibria where prices become \iid~\ In this paper we consider an averaged or fluid version of the problem and thus we only assume that the prices arrive from some common distribution. There are natural time variations in the statistics of bidding markets (due to daily and weekly cycles in human behaviour) that are not directly accounted for in these models.  Fortunately, the time varying case can largely be reduced to the time-independent case, see~\cite{kinnearbidding2_2020}, so that the results of the present paper remain applicable.

\paragraph{Outline and Contributions}
We study the optimal contract management problem for both first and second price auctions.  However, for ease of exposition, we develop the results for second price auctions first (which turn out to be slightly simpler) and then show how these results can be generalized to the first price auction, which is postponed to Section \ref{sec:first_price_case}.

In Section~\ref{sec:supply_curves} we introduce the \textit{supply curve} functions and review elementary notions for modelling the auction market.  The contract management problem with multiple contracts and multiple item types is formulated in Section~\ref{sec:contract_management}.  A special case of this problem was introduced in \cite{marbach_bidding_2020} and studied further in \cite{kinnearbidding2_2020} for non-stationary markets, where it was shown that, as long as market statistics can be accurately forecast, the non-stationary problem essentially reduces to the stationary version.

The key to the contract management problem is that it can be transformed into a \textit{convex} program.  The variables of this convex version are related to the probabilities of winning an item, rather than the bid that should be placed in the auction.  A thorough duality analysis of this convex program is given in Section~\ref{sec:contract_management_duality} where we also establish the conditions under which strong duality holds (Section~\ref{sec:contract_management_duality}).  In addition to the results shown in \cite{kinnearbidding2_2020}, we show here that consequences of duality for this problem are incredibly rich: we show that optimum bids can be recovered directly from dual multipliers, that (in certain cases) it is optimum to place the exact same bid across all item types, and that the optimum allocation decisions (from item types towards contracts) are \textit{sparse}, \ie~it is optimal to allocate only a subset of the available items towards certain contracts, depending on the average costs of these items and their value for each contract.  Convexity and convex duality also have important computational and algorithmic implications: we provide numerical example in Section~\ref{sec:computational_examples} demonstrating that large scale examples are tractable.

The main insights that enable us to study the contract management problem as a convex program appear in Section~\ref{sec:acquisition_costs}.  The key is to recognize that, while the function mapping from the bid into the average cost is merely a monotone function, the function which answers the question ``what is the expected cost to win an item with probability $q$?'' is a \textit{convex} function of $q$.  This holds under only the most elementary assumptions for second price auctions, and holds under weak assumptions on the distribution of prices in the first price case (a sufficient condition being that the distribution function of prices is log-concave).  We refer to this function as the \textit{acquisition cost function}, and its convexity enables the reliable and efficient solution of the contract management problem (some computational examples are seen in Section~\ref{sec:computational_examples}).  Details for the first price auction are provided in Section~\ref{sec:first_price_case}.  While a similar transformation appears in a proof of~\cite{fernandez2017optimal}, and the perspective of working in the quantile space, rather than with bids directly, is discussed at least as far back as~\cite{milgrom1982theory}, the convexity properties of this transformation do not appear to be widely exploited in real-time bidding (as evidenced by, to our knowledge, the lack of papers formulating optimal bidding problems as convex programs).  As well, this important property does not appear to be used or discussed by popular textbooks on auction theory~\cite{menezes2005introduction, krishna2009auction}.

The primary contributions of the present work towards the contract management problem, beyond that of~\cite{marbach_bidding_2020, kinnearbidding2_2020}, are as follows: firstly, we generalize the formulation of the contract management problem to introduce contract-dependent item valuations, which can be used to model conversion probabilities; secondly, we analyze convex duality much more thoroughly than did~\cite{kinnearbidding2_2020} in order to obtain both important structural insights and intuitive interpretations as well as to show how primal solutions can be recovered from the dual, it will be seen that convex duality also reproduces the main results of \cite{marbach_bidding_2020}, which were derived without the use of convexity; thirdly, we demonstrate that convexity and duality lead to algorithms and practical implementations for solving realistic problem instances.  We also seek to demonstrate how our techniques can apply to other related problems.  In Section~\ref{sec:related_problems} we briefly discuss the budget constrained optimum bidding problems of~\cite{zhang2014optimal} and~\cite{balseiro2015repeated}, as well as the statistical arbitrage mining problem of~\cite{zhang2015statistical}.  Once again we show how the convexity of the acquisition cost function can be used to reformulate these as convex programs, and thus we reproduce or generalize earlier results obtained without convex analysis.  Finally, we provide examples from finance, outside the field of real-time bidding: we derive an equivalence between the acquisition cost function in second price auctions and the cost of submitting market orders in limit order book financial markets; we use this insight to model, as a convex program, a novel portfolio construction problem; we also show a connection with the cost function arising in the dark pool problem of~\cite{ganchev2010censored}.  Section~\ref{sec:conclusion} concludes the paper.

\section{Cost Functions for First and Second Price Auctions}
\label{sec:supply_curves}
In this section, we introduce the basic concepts essential for formulating the contract management problem.  We formally define the fundamental notion of a \textit{supply curve} in Section~\ref{sec:supply_curves_subsec}, and the expected cost functions of bidding in second price auctions in Section~\ref{sec:cost_functions}.  For ease of exposition, we defer the consideration of first price auctions, which require only modest modifications, to Section \ref{sec:first_price_case}.

\begin{figure}
  \centering
  \begin{subfigure}[b]{0.4\textwidth}
    \includegraphics[width=\textwidth]{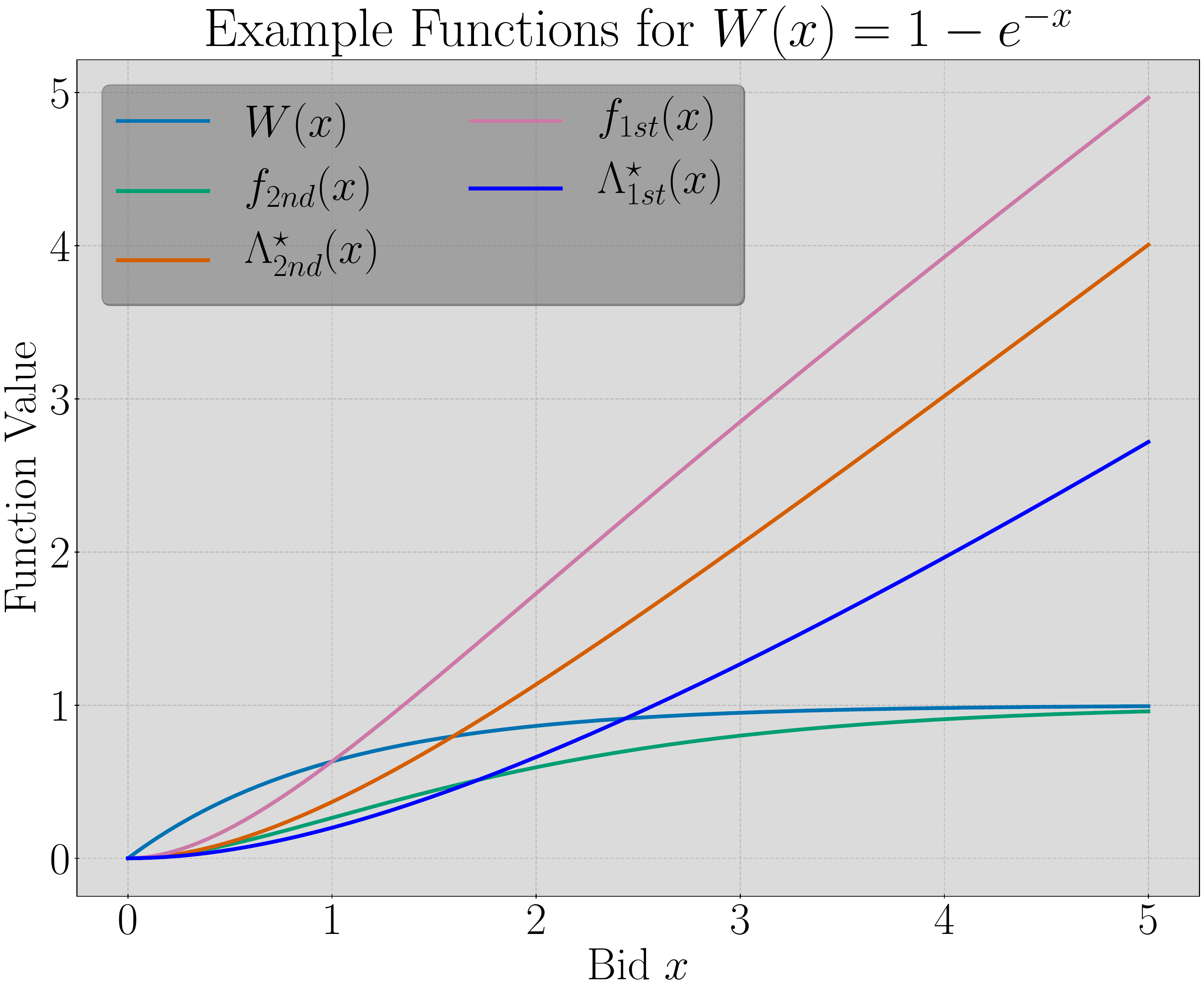}
    \caption{Functions of the Bid \texorpdfstring{$x$}{x}}%
    \label{fig:example_functions_primal}
  \end{subfigure}
  \begin{subfigure}[b]{0.4\textwidth}
    \includegraphics[width=\textwidth]{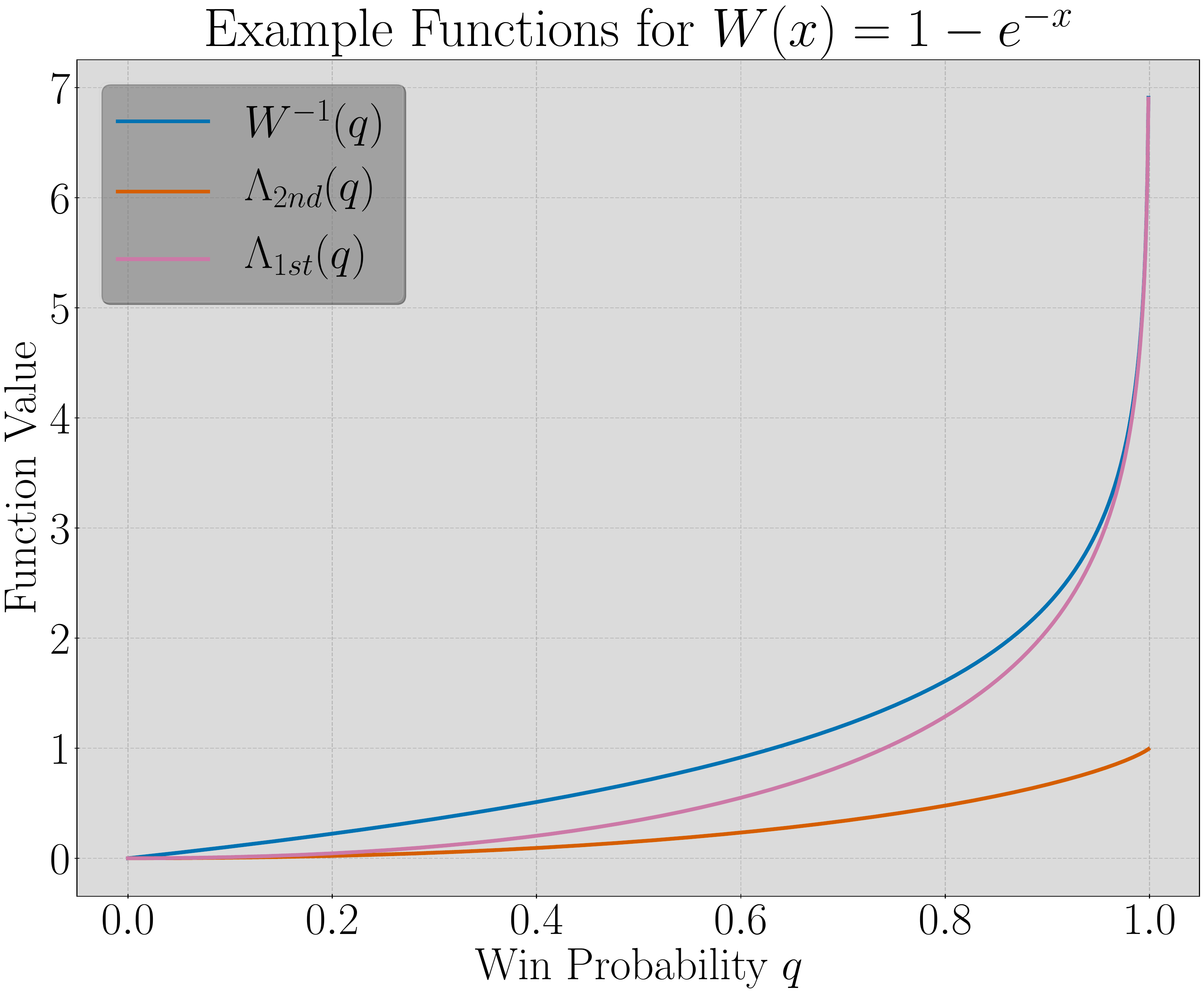}
    \caption{Functions of Win Probability \texorpdfstring{$q$}{q}}%
    \label{fig:example_functions_dual}
  \end{subfigure}

  {\small Important example functions for the simple case that $W(x) = 1 - e^{-x}$, \ie~where the price is $\text{exp}(1)$ distributed.  These functions are introduced as needed throughout the paper and the figure can be referenced whenever they are first encountered.}%
  \centering
  \caption{Illustrative Example Functions}%
  \label{fig:example_functions}
\end{figure}

We provide early on an illustrative example of important functions encountered in this paper in Figure~\ref{fig:example_functions}.  This figure can be referenced throughout as the functions $W(x)$ (the supply curve), $f(x)$ (the cost curve), $\Lambda(q)$ (the acquisition cost function), $\Lambda^\star(\mu)$ (the dual acquisition cost function) are introduced.

\subsection{Supply Curves}%
\label{sec:supply_curves_subsec}
Consider an auction market with indistinguishable \textit{items} (to use generic terminology, more specifically, the items are impressions).  Throughout, the discussion will generally be from the perspective of a particular DSP placing a bid (or bids) $x$ for an item that becomes available in the auction.  The \textit{supply curve}, which is a cumulative distribution function (\cdf), characterizes the bidder's competition by quantifying the probability of winning an item given that the bidder places the bid $x$ (see~\cite{friedman1956competitive} for some of the earliest work in auction theory using this idea).  This function will be denoted generically as $W: \R \rightarrow [0, 1]$ and signifies that the probability of winning an item with the bid $x$ is $W(x)$.

We assume that the \textit{price} $p$ (which is random and hence unknown to the bidder), of an item arriving at the auction is distributed according to the \cdf~$W$, and that the item will be won by placing a bid $x \ge p$ as in $W(x) = \P\{p \le x\}$.  Additionally, we will assume that prices are sequentially independent and that items arrive at an average rate $\lambda > 0$; thus, the average rate that items will be won by placing the bid $x$ on every arriving item is simply $\lambda W(x)$.  The following definition formally specifies the properties that $W$ must have.  

\begin{definition}[Supply Curve and Market Model]%
  \label{def:supply_curve}
  A supply curve $W(x)$ is a continuous cumulative distribution function on $\R$ such that $\forall x \le 0:\ W(x) = 0$, and that $W$ is strictly monotone from $x = 0$ up to some maximum bid $\bar{x}$ which may be $\infty$. If $\bar{x} = \infty$, we extend $W$ by continuity so that $W(\infty) \defeq \underset{x \rightarrow \infty}{\text{lim}}\ W(x) = 1$. The first moment $\bar{p} \defeq \int_0^\infty u\d W(u) < \infty$ is finite.  The prices (\ie~the highest competing bids) $p_1, p_2, \ldots$ of arriving items are distributed \textit{i.i.d.} according to $W$ such that $\P\{p_n \le x\} = W(x)$ models the probability of winning an item given a bid of $x$.  The inverse of $W$ on $[0, 1]$ is denoted $W^{-1}(q)$ and has $W^{-1}(1) = \bar{x}$.  The inverse is extended such that $W^{-1}(q) = 0$ if $q \le 0$ and $W^{-1}(q) = \infty$ if $q > 1$.
\end{definition}

\begin{remark}
  In contrast to ~\cite{marbach_bidding_2020}, where $W$ was allowed to be an arbitrary \cdf, We assume that $W$ is continuous, and strictly increasing.  This additional regularity is essential for our main results.  We can remark that if the DSP estimates $W$ through a kernel density estimate (a natural estimator to use for $W$) then the corresponding $W(.)$ will be completely smooth.
\end{remark}
\subsection{Cost of Bidding}%
\label{sec:cost_functions}
We will denote by $f(x)$ the expected cost of bidding $x$ on an item of some fixed type.  This function depends upon the particular auction mechanism that is employed by the platform, as well as the supply curve $W(x)$ associated with that item type.  The most common mechanisms are \textit{first price} and \textit{second price} auctions.  In the former case, the winner of the auction pays what they bid, and in the latter, the winner pays the price that reflects the highest competing bid.  Additionally, the seller of the item may incorporate a \textit{reserve price} for the auction which is such that any bid below the reserve price is immediately rejected.  Reserve prices can arise if the seller has a personal valuation for the item (\eg~a website may associate some value with advertising their own premium service to their own users) or the seller may have alternative platforms where they are guaranteed to be able to sell their items above some specified price.  Since there are a large number of sellers, and the sellers may have additional private and independent information about the items that affects their personal value of the items, we treat the reserve price as though it arises from another competing bidder, and do not consider it separately.

It will be seen in Section \ref{sec:first_price_case} that the analysis of both the first and second price auctions can be unified.  Thus, we will generically write $f$ for a cost function when the auction type turns out to be irrelevant, or is specified in the context.  But, when it is necessary, we will notationally distinguish the costs in first and second price auctions with $f_{1st}$ or $f_{2nd}$, respectively.  In a second price auction, we have

\begin{equation}
\label{eqn:second_price_cost}
  f_{2nd}(x) = \E[p \ind(p \le x)] = \int_0^x u\d W(u).
\end{equation}
where $\ind(P) $ the indicator function for a proposition $P$. A similar cost function is obtained in the context of limit order books and discussed in Section~\ref{sec:lob_problem}.  The first price case is deferred to Section~\ref{sec:first_price_case}.

\section{Optimal Contract Management in Real-time Bidding}
\label{sec:contract_management}
We turn now to the formulation and solution of the main problem.  Suppose that there is a finite number of \textit{item types} $j \in [M] \defeq \{1, 2, \ldots, M\}$, each with their own associated supply curves, cost functions,~\etc\ all of which are indexed by $j$ as in $W_j(x)$,~\etc\  Items of type $j$ are assumed to arrive at the auction at the average rate $\lambda_j > 0$ so that the average cost of bidding $x$ on every arriving item of type $j$ is $\lambda_j f_j(x)$ and the average number of items won with such bids is given by $\lambda_j W_j(x)$.

\begin{remark}[Notation]
  In order to keep notation light, the subscript $j$ will be omitted from $W_j, f_j$~\etc whenever we are discussing a single fixed item type.  Subscripts $2nd$ and $1st$ are sometimes omitted as well if it has been specified which auction is being discussed (or if the type of auction turns out to be irrelevant).  As well, the constraints in optimization problems appear in this paper are to be understood to hold over all unquantified indices.  For example, a constraint $\sum_{i = 1}^N v_{ij} = x_j$, should be understood to mean $\sum_{i = 1}^N v_{ij} = x_j\ \forall j \in [M].$
\end{remark}

We are interested in a \textit{contract management} problem, a version of which was first introduced, without differing item valuations, by~\cite{marbach_bidding_2020}, and further studied by~\cite{kinnearbidding2_2020}.  A \textit{contract} in this context, of which there will be $N$, is a tuple $\bigl(C_i, {(v_{ij})}_{j \in [M]}\bigr)$ where $v_{ij}$ is the value of items of type $j$ towards contract $i \in [N]$, and $C_i > 0$ is a target rate at which items are to be obtained.  The valuations $v_{ij} \in \R_+$ can be interpreted as click through rates, but it is not formally necessary to assume that they are bounded by $1$.  Formally, the contract management problem is described by:

\begin{equation}
  \label{eqn:main_problem_Wf}
  \begin{aligned}
    \underset{x, \gamma}{\text{minimize}} &\quad \sum_{j = 1}^M \sum_{i = 1}^N \gamma_{ij} f_j(x_{ij})\\
    \text{subject to}
    &\quad \sum_{j = 1}^M \gamma_{ij} \lambda_j v_{ij} W_j(x_{ij}) = C_i\\
    &\quad \sum_{i = 1}^N \gamma_{ij} \le 1, \gamma_{ij} \ge 0,
  \end{aligned}\tag{$P^m$}
\end{equation}
where $x_{ij} \in \R$ is the bid to be placed on items of type $j$ when they will be allocated towards contract $i$, and $\gamma_{ij} \in \R$ is the proportion of items of type $j$ which, when won, should be allocated towards the fulfillment of contract $i$.

\begin{remark}[Implementation and Interpretation]
  Given a solution $\gamma, x$ to Problem \eqref{eqn:main_problem_Wf}, it is used in the RTB market as follows: whenever an item of type $j$ arrives, select contract $i \in [N]$ with probability $\gamma_{ij}$ and place the bid $x_{ij}$.  If the item is won, allocate it towards the fulfillment of contract $i$.
\end{remark}

\begin{definition}[Sparsity Inducing Sets $\A_i, \B_j$]
  Problem~\eqref{eqn:main_problem_Wf} nominally has $\mathcal{O}(MN)$ variables and constraints, but in many practical cases it is likely that this can be dramatically reduced.  Typically, the number of item types $M$ will be large and satisfy $M \gg N$, and the valuations $v_{ij}$ should be sparse in the sense that $v_{ij} = 0$ for most $i, j$.  That is, only a small collection $\A_i \defeq \{j \in [M]\ |\ v_{ij} > 0\}$ of item types are actually useful for contract $i$.  Likewise, each item type will only be useful for fulfilling a select number of contracts $\B_j \defeq \{i \in [N]\ |\ v_{ij} > 0\}$.  If we denote by $d_i = |\A_i|$ the cardinality of the set $\A_i$ and $d = \sum_{i = 1}^N d_i$ as the total dimension, then Problem~\eqref{eqn:main_problem} can be modified to have only $\mathcal{O}(d)$ variables and constraints by restricting the summations to only $\sum_{j \in \A_i} \bullet$ and $\sum_{i \in \B_j} \bullet$~\etc\  Thus, one should think of $x_{ij}, \gamma_{ij}$ as being arrays which are defined only for $v_{ij} > 0$, rather than as dense matrices in $\R^{M \times N}$.
\end{definition}

\subsection{Convexity}%
\label{sec:acquisition_costs}
Problem \eqref{eqn:main_problem_Wf} is a monotone programming problem in the sense of~\cite{tuy2000monotonic}, but such a problem is still not tractable.  That is, while there are convergent algorithms for solving such problems, the rate at which these algorithms converge is slow.  However, via a transformation of variables, we can show that Problem \eqref{eqn:main_problem_Wf} is equivalent to a \textit{convex} optimization problem~\eqref{eqn:main_problem}.  We carry out this transformation in the present section.

Since any supply curve $W$ is strictly monotone over an interval $[0, \bar{x}) \subseteq [0, \infty)$ and hence is one-to-one, we can calculate what bid is needed to acquire the item with probability $q \in [0, 1]$, that is, $x = W^{-1}(q)$, with $W^{-1}(q) \defeq \infty$ whenever $\bar{x} = \infty$.  It is then natural to ask a related question: ``what is the expected cost to win the item with probability $q$?''.  The answer to this question is the \textit{acquisition cost function}

\begin{equation}
  \label{eqn:acquisition_cost}
  \Lambda(q) = f \circ W^{-1}(q).
\end{equation}
By direct substitution of $W^{-1}(q)$ into $f_{2nd}$ we can see that, for $q \in [0, 1]$:

\begin{equation}
  \label{eqn:acquisition_cost_auction}
  \begin{aligned}
    \Lambda_{2nd}(q) &= \int_0^{W^{-1}(q)} u\d W(u).
  \end{aligned}
\end{equation}
In order to extend the definition of $\Lambda$ to all of $\R$, we let $\Lambda(q) = \infty$ on $q > 1$ and $\Lambda(q) = 0$ for $q < 0$. The remarkable property of this function is that it is \textit{convex}.

\begin{theoremEnd}[normal]{proposition}[Convex Acquisition Costs --- Second Price Case]%
  \label{prop:convex_acquisition_costs_2pa}
  Let $W(x)$ be a supply curve.  Then, in a second price auction, the acquisition cost function $\Lambda_{2nd}(q) = f_{2nd} \circ W^{-1}(q)$ is given by $\int_0^q W^{-1}(u)\d u$ on $q \in [0, 1]$.  If this is extended to:

  \begin{equation}
    \label{eqn:integral_acq_costs_2pa}
    \Lambda_{2nd}(q) \defeq \left\{\begin{array}{lr}
                               \infty; & q > 1\\
                               0; & q \le 0\\
                               \int_0^q W^{-1}(u)\d u; & \mathrm{otherwise}
                             \end{array}\right.,
  \end{equation}
then $\Lambda_{2nd}$ is a proper (recall that a function $f: \R \rightarrow (-\infty, \infty]$ is \textit{proper} if it is not everywhere equal to $+\infty$), lower semi-continuous, non-decreasing, and convex function on $\R$.  Moreover, $\Lambda_{2nd}$ is strictly convex over $[0, 1]$, differentiable on $(0, 1)$, and the derivative can be extended continuously to $[0, 1]$ if $\bar{x} < \infty$.

% It's obvious that \Lambda(q) = \E[p]$ ?
\end{theoremEnd}
\begin{proofEnd}
\textit{Proof.}\ We first calculate $\Lambda = f_{2nd} \circ W^{-1}$ for $q \in [0, 1]$ by

  \begin{align*}
    \Lambda(q)
    &\overset{(a)}{=} \int_0^{W^{-1}(q)} x \d W(x)\\
    &\overset{(b)}{=} \int_0^q W^{-1}(u)\d u,
  \end{align*}
  where $(a)$ follows by definition of $f_{2nd}$ (Equation~\eqref{eqn:second_price_cost}) and $(b)$ is by the change of variables $v = W(x)$.  We see from this latter formula that $\Lambda_{2nd}$ is differentiable on $(0, 1)$ with derivative $W^{-1}(q)$, which is continuous over $[0, 1]$ when $\bar{x} < \infty$.  Since $W$ is strictly monotone, $W^{-1}$ is also strictly monotone, and functions with strictly monotone derivatives are strictly convex, it follows that $\Lambda_{2nd}$ is strictly convex over $(0, 1)$.  The extension as given maintains convexity, and ensures that $\Lambda_{2nd}$ is lower semi-continuous.  It is proper since, \eg~$\Lambda(1) = \int_0^\infty u\d W(u) < \infty$.
\end{proofEnd}

% LEMMA HERE MOVED TO APPENDIX

Using the convexity of the function $\Lambda(q)$, we can reformulate Problem~\eqref{eqn:main_problem_Wf} into an equivalent and tractable convex program.

\begin{theoremEnd}[normal]{proposition}[Convex Reformulation]
  \label{prop:convex_reformulation}
  In a second price auction, Problem~\eqref{eqn:main_problem_Wf} can be reformulated as

  \begin{equation}
    \label{eqn:main_problem}
    \begin{aligned}
      \underset{s, R}{\mathrm{minimize}} &\quad \sum_{j = 1}^M \lambda_j \Lambda_j(s_j / \lambda_j)\\
      \mathrm{subject\; to}
      &\quad \sum_{j \in \A_i} v_{ij} R_{ij} = C_i\\
      &\quad \sum_{i \in \B_j} R_{ij} = s_j, R_{ij} \ge 0.
    \end{aligned}
    \tag{$P$}
  \end{equation}
    If a solution exists, then a solution to the original problem~\eqref{eqn:main_problem_Wf} is obtained via $x_{ij} = W_j^{-1}(s_j / \lambda_j)$ and $\gamma_{ij} = R_{ij} / s_j$ (with $0 / 0 \defeq 0$)  for each $i \in [N], j \in \A_i$.  Moreover, Problem~\eqref{eqn:main_problem} is a convex optimization problem (in the sense of~\cite{boyd2004convex}), since it has a convex objective function and affine (hence convex) constraints.
\end{theoremEnd}
\begin{proofEnd}
  \textit{Proof.}\ We apply Lemma~\ref{lem:ubp} to first eliminate the dependence of the bid on $i$, since if a solution exists it can be assumed to have the property $x_{ij} = x_j$:

    \begin{equation*}
      \begin{aligned}
        \underset{x, \gamma}{\text{minimize}} &\quad \sum_{i = 1}^N \sum_{j = 1}^M\gamma_{ij}\lambda_jf_j(x_j)\\
        \textrm{subject to}
        &\quad \sum_{j = 1}^M\gamma_{ij}\lambda_jv_{ij}W_j(x_j) = C_i\\
        &\quad \sum_{i = 1}^N \gamma_{ij} \le 1, \gamma_{ij} \ge 0,
      \end{aligned}
    \end{equation*}
    Since the bids do not depend on $i$, we can rearrange the objective
    by swapping the order of summation.  That is, at an optimal solution:

    \begin{equation}
      \sum_{i = 1}^N \sum_{j = 1}^M\gamma_{ij}\lambda_jf_j(x_j)
      = \sum_{j = 1}^M\lambda_jf_j(x_j) \sum_{i = 1}^N\gamma_{ij}
      \overset{(a)}{=} \sum_{j = 1}^M\lambda_jf_j(x_j),
    \end{equation}
    where $(a)$ follows since $\sum_{i = 1}^N\gamma_{ij} \in \{0, 1\}$ (see Lemma~\ref{lem:ubp}), and if $\sum_{i = 1}^N\gamma_{ij} = 0$ we can assume without loss that $x_j = 0$ since $f_j(0) = 0$.  Then, making the substitution $s_j = \lambda_j W_j(x_j)$, results in
    \begin{equation}
      \begin{aligned}
        \underset{s, \gamma}{\text{minimize}} &\quad \sum_{j = 1}^M\lambda_j\Lambda_j(s_j / \lambda_j)\\
        \textrm{subject to}
        &\quad \sum_{j = 1}^M\gamma_{ij}v_{ij}s_j \ge C_i\\
        &\quad \sum_{i = 1}^N \gamma_{ij} = 1\\
        &\quad \gamma_{ij} \ge 0, s_j \ge 0.
      \end{aligned}
    \end{equation}
    Finally,  substituting $R_{ij} = \gamma_{ij}s_j$, and eliminating the variable $\gamma$, results in the formulation of Problem~\eqref{eqn:main_problem}.  That $\gamma_{ij}$ can be dropped from the problem follows from the fact that the objective function is independent of $\gamma$, and therefore, for any value of $s, R$, a feasible $\gamma$ can be constructed which satisfies its constraints and $R_{ij} = \gamma_{ij}s_j$ without affecting the objective function value.  Problem~\eqref{eqn:main_problem} is a convex optimization problem because the objective function is convex, the inequality constraints are specified by a finite number of convex (in fact, affine) functions, and the equality constraints are affine.
\end{proofEnd}

In Problem~\eqref{eqn:main_problem}, the variable $s_j$ is the rate at which items of type $j$ are to be acquired, and $R_{ij}$ is the rate at which items of type $j$ are to be allocated towards the fulfillment of contract $i$.  The optimal bids $x_{ij} = W_j^{-1}(s_j / \lambda_j)$ are such that the right hand side of this equation do not depend upon the contract, thus, we need only consider bids $x_j$: the optimal bid does not depend upon the contract towards which the item will be allocated.  The transformation of variables applied to Problem~\eqref{eqn:main_problem_Wf} can be inferred from here: we let $s_j = \lambda_j W_j(x_j)$ and $R_{ij} = \gamma_{ij} s_j$.  We remark that the substitution $R_{ij} = \gamma_{ij}s_j$ may appear similar to the technique of \textit{linearization} for relaxing intractable problems into convex problems.  We emphasize that in the case studied here, there is an exact correspondence, not a relaxation, between \eqref{eqn:main_problem_Wf} and \eqref{eqn:main_problem}.

\subsection{Duality: Consequences and Interpretations}%
\label{sec:contract_management_duality}

Before proceeding, we need to make a basic assumption that will guarantee existence of solutions to Problem~\eqref{eqn:main_problem}:

\begin{assumption}[Adequate Supply]%
  \label{ass:adequate_supply}
  There exists an array $R_{ij}$ of allocations such that $\forall i \in [N], j \in \A_i:\ R_{ij} \ge 0$, $\forall j \in [M]:\ \sum_{i \in \B_j} R_{ij} < \lambda_j$, and $\forall i \in [N]:\ \sum_{j \in \A_i} v_{ij} R_{ij} = C_i$.
\end{assumption}

This assumption is natural in the sense that if it were not satisfied, the contracts would not be able to be fulfilled, and they should not have been accepted in the first place.  Essentially, if Assumption~\ref{ass:adequate_supply} holds, then Problem~\eqref{eqn:main_problem} admits a \textit{Slater Point} (see \eg~\cite[Sec. 5.2.3]{boyd2004convex}~\cite[Prop 5.3.1]{bertsekas2009convex}); the inequality must be strict since the domain of $\Lambda_j$ implicitly imposes the constraint that $s_j \le \lambda_j$, and this property is later used to guarantee that the bids required to obtain the optimal supply rates are finite.  An assumption similar in spirit, but much stronger, appears in~\cite{leblanc1974transportation}.

We now derive a dual of Problem~\eqref{eqn:main_problem} and then proceed to an analysis and interpretation of the dual.  We begin with the Lagrangian function associated to Problem~\eqref{eqn:main_problem}:

\begin{align*}
\L(s, R, \mu, \rho, \theta)
&= \sum_{j = 1}^M \lambda_j \Lambda_j \bigl(s_j / \lambda_j \bigr) + \sum_{i = 1}^N \rho_i \bigl(C_i - \sum_{j \in \A_i}R_{ij}v_{ij}\bigr) + \sum_{j = 1}^M \mu_j\bigl(\sum_{i \in \B_j} R_{ij} - s_j\bigr) - \sum_{j = 1}^M\sum_{i \in \B_j} \theta_{ij}R_{ij}\\
&= \sum_{i = 1}^N \rho_i C_i + \sum_{j = 1}^M\Bigl[\lambda_j \Lambda_j \bigl(s_j / \lambda_j\bigr) - \mu_j s_j + \sum_{i \in \B_j} R_{ij} \bigl(\mu_j - \theta_{ij} - v_{ij}\rho_i\bigr) \Bigr],
\end{align*}
where $\mu \in \R^M$ is associated with the equality constraints $\sum_{i \in \B_j} R_{ij} = s_j$, $\rho \in \R^N$ is associated with the contract fulfillment constraints, and $\theta$ to the non-negativity constraints.  The dual constraints are $\theta \ge 0$ and the dual problem is derived through determining the form of \[\underset{\rho, \mu, \theta \ge 0}{\text{maximize}}\ \underset{s, R}{\text{inf}}\ \L(s, R, \mu, \rho, \theta).\]  To this end,  we minimize $\L$ pointwise over $s$, which results in the appearance of the Fenchel conjugate~\cite[Sec 4.2]{clarke2013functional}

\begin{equation*}
  \Lambda^\star(\mu) \defeq  \text{sup}_q\ [\mu q - \Lambda_{2nd}(q)],
\end{equation*}
that is, minimizing $\L$ over $s$ we obtain:

\begin{equation}
  \label{eqn:lagrangian_s}
  \underset{s}{\text{inf}}\ \L(s, R, \mu, \rho, \theta) = \sum_{i = 1}^N \rho_i C_i - \sum_{j = 1}^M \lambda_j \Lambda_j^\star(\mu_j) + \sum_{j = 1}^M\sum_{i \in \B_j} R_{ij} \bigl(\mu_j - \theta_{ij} - v_{ij}\rho_i \bigr).
\end{equation}
There is an appealing duality relationship between $\Lambda_{2nd}$ and $\Lambda_{2nd}^\star(\mu)$, namely that $\Lambda_{2nd}$ is the integral of $W^{-1}$ and that $\Lambda_{2nd}^\star$ is the integral of $W$, see~\cite{gushchin2017integrated} for further analysis of integrated quantile functions.

\begin{theoremEnd}[normal]{proposition}[Fenchel Conjugate --- Second Price Case]%
  \label{prop:acquisition_cost_duality}
  The Fenchel duality relationship between $\Lambda_{2nd}$ and $\Lambda_{2nd}^\star$ is between the integrated c.d.f. $W$ and the integrated quantile function $W^{-1}$:

  \begin{equation}
    \Lambda_{2nd}(q) \defeq \left\{\begin{array}{lr}
                               \infty; & q > 1\\
                               0; & q \le 0\\
                               \int_0^q W^{-1}(u)\d u; & \mathrm{otherwise}
                             \end{array}\right.,
  \end{equation}

  \begin{equation}
  \label{eqn:2pa_conjugate}
    \Lambda_{2nd}^\star(\mu) =
     \begin{cases}
       \infty, & \text{if } \mu < 0\\
       \int_0^\mu W(u)\d u, & \text{if } \mu \in [0, \bar{x}]\\
       \mu - \bar{x}, & \text{if } \mu > \bar{x}.
     \end{cases}
  \end{equation}
  The function $\Lambda_{2nd}^\star$ is a proper, convex, and lower-semicontinuous function, which is strictly convex and strictly monotone increasing $\R_+$.
\end{theoremEnd}
\begin{proofEnd}
  \textit{Proof.}\ By definition, \[\Lambda_{2nd}^\star(\mu) = \underset{q \in (-\infty, 1]}{\text{sup}}\bigl[\mu q - \Lambda_{2nd}(q)\bigr],\] where the domain is restricted to $(-\infty, 1]$ since $\Lambda_{2nd}(q) = \infty$ for $q > 1$.  If $\mu < 0$ then $\Lambda_{2nd}^\star(\mu) = \infty$ since $\Lambda_{2nd}(q) = 0$ for $q \le 0$.  If $\mu > \bar{x}$ (note that this case is excluded if $\bar{x} = \infty$ since then $\mu > \bar{x}$ cannot be) then $\Lambda^\star(\mu) = \mu - \bar{x}$ since $\mu$ is greater than the maximal slope of $\Lambda$.  Finally, if $\mu \in [0, \bar{x}]$ then we can solve by differentiation to obtain that, optimally, $q = W(\mu)$ and hence

  \begin{align*}
    \Lambda_{2nd}^\star(\mu)
    &= \mu W(\mu) - \Lambda_{2nd} \circ W(\mu)\\
    &= f_{1st}(\mu) - f_{2nd}(\mu)\\
    &= \int_0^\mu W(u)\d u,
  \end{align*}
  where the final equality follows through integration by parts.  It is proper, convex, and lower-semicontinuous since it is a conjugate function, and strictly convex on $\R_+$ since $W$ is strictly monotone on $[0, \bar{x}]$ and the function is linear thereafter.  It remains strictly convex on the rest of $\R$ since it is affine for $\mu > \bar{x}$.  It is strictly monotone on $\R_+$ by inspection.
\end{proofEnd}

\begin{remark}[Derivatives --- Second Price Case]%
  \label{rem:derivatives}
  From Propositions~\ref{prop:convex_acquisition_costs_2pa} and~\ref{prop:acquisition_cost_duality} we can conclude that for $q \in (0, 1)$ the derivative of the acquisition cost $\Lambda_{2nd}(q)$ is given by the inverse of the supply curve $W^{-1}(q)$ and the derivative of the conjugate $\Lambda^\star_{2nd}(\mu)$ is the supply curve itself $W(\mu)$.  Therefore, in many cases it may be easier to solve dual optimization problems involving $\Lambda^\star$, rather than primal problems involving $\Lambda$, since the derivatives of the former may be more easily available.  These properties can be applied to derive very simple stochastic approximation algorithms to learn optimum bids without direct knowledge of $W$; these results will be reported elsewhere, see also the thesis \cite{kinnearphdthesis}.
\end{remark}

Returning to the Lagrangian $\L$ and minimizing over $R$ induces the dual equality constraint $\mu_j = \theta_{ij} + v_{ij}\rho_i $ (since otherwise the infimum would be $-\infty$), which must hold over all $(i, j)$ such that $i \in [N], j \in \A_i$, or equivalently, $j \in [M], i \in \B_j$.  Moreover, since we must have the dual cone constraint $\theta_{ij} \ge 0$ (\ie~$\theta_{ij}$ is simply a slack variable) this is equivalent to the constraint $\forall i \in [N], j \in \A_i:\ \mu_j \ge \rho_i v_{ij}$.  Thus, we have obtained:

\begin{theoremEnd}[normal]{proposition}[Duality]
  A dual of Problem~\eqref{eqn:main_problem} can be formulated as 

  \begin{equation}
    \label{eqn:dual_problem}
    \begin{aligned}
    \underset{\rho, \mu}{\text{maximize}} &\quad \sum_{i = 1}^N \rho_i C_i - \sum_{j = 1}^M \lambda_j \Lambda_j^\star (\mu_j)\\
    \text{subject to}
    &\quad v_{ij}\rho_i \le \mu_j,
    \end{aligned}\tag{D}
  \end{equation}
  where the linear inequality constraints hold over all $i \in [N], j \in \A_i$.  Problem~\eqref{eqn:dual_problem} is dual to Problem~\eqref{eqn:main_problem} in the sense that if $D^\star$ and $P^\star$ are their respective values (possibly $\infty$ or $-\infty$), then $D^\star \le P^\star$.
\end{theoremEnd}

Problem~\eqref{eqn:dual_problem} has $N + M$ variables, and $d = \sum_{i = 1}^N |\A_j|$ linear inequality constraints.  This is in contrast to the primal~\eqref{eqn:main_problem} which has $d$ variables, $N + M$ linear equality constraints and $d$ non-negativity constraints.  In Section~\ref{sec:consequences_of_duality} we show how to recover primal solutions from dual solutions.

\begin{remark}[Implicit Constraints and Non-negativity]
  In the primal, Problem~\eqref{eqn:main_problem}, there is an implicit constraint $s_j \le \lambda_j$ due to the domain of $\Lambda_j$.  Similarly, in the dual~\eqref{eqn:dual_problem} there
is an implicit constraint that $\mu_j \ge 0$ due to the domain of $\Lambda_j^\star$.  Additionally, as a consequence of the monotonicity of the objective (\ie~$\rho_i C_i$ is strictly monotone increasing since $C_i > 0$), it is easy to observe that at the optimal solution the dual variables $\rho_i$ will also be non-negative: $\rho \ge 0$.  It is not formally necessary to include these as constraints in~\eqref{eqn:dual_problem}, but we have observed improved numerical optimization performance (with \texttt{cvxpy}~\cite{diamond2016cvxpy}) when they are made explicit, particularly for large problem instances (1000s of contracts and/or item types).
\end{remark}

Under Assumption~\ref{ass:adequate_supply}, at least one solution $(s, R)$ to Problem~\eqref{eqn:main_problem} exists, strong duality holds, and the bids $x_j$ such that $\lambda_j W_j(x_j) = s_j$ are finite.

\begin{theoremEnd}[normal]{proposition}[Existence and Regularity]%
  \label{prop:regularity}
  Suppose Assumption~\ref{ass:adequate_supply} holds.  Then, there exists an optimal allocation array $R^\star$ for Problem~\eqref{eqn:main_problem}, the optimal win rates $s^\star_j = \sum_{i \in \B_j} R_{ij}^\star$ are unique and there exist unique finite bids $x_j = W_j^{-1}(s_j / \lambda_j)$ which win items at the optimal rate.  Moreover, there exists a unique solution $\rho^\star, \mu^\star$ to the dual, Problem~\eqref{eqn:dual_problem}, and there is zero duality gap.
\end{theoremEnd}
\begin{proofEnd}
  \textit{Proof.}\ By Assumption~\ref{ass:adequate_supply}, the value of Problem~\eqref{eqn:main_problem} is finite since there exists a feasible $s, R$.  Then, since $\Lambda_j$ is lower semicontinuous and the feasible region is compact, namely $R_{ij} \ge 0$ and $\sum_{i \in \B_j} R_{ij} \le \lambda_j$ (which is implicit in the domain of $\Lambda_j$), there exists a solution $s^\star, R^\star$ to Problem~\eqref{eqn:main_problem} by Weierstrass' theorem.  The acquisition rates $s_j^\star = \sum_{i \in \B_j} R_{ij}^\star$ are unique since $\Lambda_j$ is strictly convex.

Now, by the strong duality theorem (\eg~\cite[Prop 5.3.1]{bertsekas2009convex}) and the Slater point of Assumption~\ref{ass:adequate_supply}, there exists a solution $\mu^\star, \rho^\star$ (along with the slack variables $\theta^\star$) to the dual, Problem~\eqref{eqn:dual_problem}, and the value of the dual, which is finite, is equal to the value of the primal (\ie~there is zero duality gap).

Finally, by strong duality and the existence of solutions, the Lagrangian optimality conditions (\eg~\cite[Prop 5.3.2]{bertsekas2009convex}) establish that $s^\star \in \underset{s \in \R^M}{\text{argmin}}\ \L(s,R^\star, \mu^\star, \rho^\star, \theta^\star)$ and \[\L(s, R^\star, \mu^\star, \rho^\star, \theta^\star) = \sum_{i = 1}^N \rho_i^\star C_i + \sum_{j = 1}^M [\lambda_j \Lambda_j(s_j / \lambda_j) - \mu^\star_j s_j],\] where we have used complementary slackness~\cite[prop 5.3.2]{bertsekas2009convex} to eliminate terms involving $\theta^\star, R^\star$.  Then, by the subgradient optimality conditions~\cite[Prop. 5.4.3, Prop 5.4.4]{bertsekas2009convex}, it must be that $\forall j \in [M]:\ 0 \in \partial_{s_j} \L(s^\star, R^\star, \mu^\star, \rho^\star, \theta^\star) = \lambda_j \partial \Lambda_j(s_j^\star / \lambda_j) - \mu_j^\star$.  We treat two cases.  First, if the support of the distribution of the prices of type $j$ is compact (\ie~$\bar{x}_j < \infty$), then since $s_j \le \lambda_j$ we have a finite optimal bid $x_j = W_j^{-1}(s_j / \lambda_j)$.  If $\bar{x}_j = \infty$ then $\partial \Lambda_j(q)$ is non-empty for $q \in [0, 1)$ and thus, since we know this subgradient is indeed non-empty, we must have $s_j < \lambda_j$ and there is a unique optimal bid $x_j = W_j^{-1}(s_j / \lambda) < \infty$.
\end{proofEnd}

\subsubsection{Consequences of Duality}%
\label{sec:consequences_of_duality}
There are three main consequences of duality.  Firstly, the dual variables induce subsets $\A_i^\star \subseteq \A_i$ and $\B_j^\star \subseteq \B_j$ which have the effect of further reducing the set of useful and usable items.  That is, for an optimal allocation array $R$, it holds that $j \not\in \A_i^\star \implies R_{ij} = 0$.  These sets considerably reduce the dimensionality of the allocation array $R$ that needs to be calculated, if dual solutions are known.  Secondly, the optimal bids to be placed are (in second price auctions) exactly equal to the dual multipliers: $x_j = \mu_j$.  These optimum bids can be characterized either directly by variables $\mu_j$ associated to each item type, or indirectly through the vector of $N$ \textit{pseudo-bids}~\cite{kinnearbidding2_2020} $\rho \in \R^N$, with each $\rho_i$ being associated with a contract, rather than an item type.  Moreover, the optimal bid to be placed across all items belonging to the set $\A_i^\star$ can be obtained through the relation $j \in \A_i^\star \implies \mu_j = v_{ij} \rho_i$.  Finally, an optimal allocation matrix $R$ can be recovered from the dual variables (\ie~from the double dual) of Problem~\eqref{eqn:dual_problem}.

We will summarize these facts in Proposition~\ref{prop:duality_consequences}.  These results reproduce many of which were derived directly, rather than as consequences of duality, by~\cite{marbach_bidding_2020}.  We summarize these results in the following proposition.

\begin{theoremEnd}[normal]{proposition}[Consequences of Duality]%
  \label{prop:duality_consequences}
  Suppose Assumption~\ref{ass:adequate_supply} holds.  Let $R \in \R^{N \times M}$ and $s \in \R^M$ be optimal primal solutions of Problem~\eqref{eqn:main_problem} such that $\sum_{i \in \B_j} R_{ij} = s_j$ and $\sum_{j \in \A_i}^M R_{ij} v_{ij} = C_i$.  As well, Let $\rho \in \R^N$ and $\mu \in \R^M$ constitute optimal dual solutions to Problem~\eqref{eqn:dual_problem} and let $\theta_{ij} = \mu_j - v_{ij} \rho_i$ be the slack for $i \in [N], j \in \A_i$.  Finally, let $x_j = W_j^{-1}(s_j / \lambda_j)$ be the optimal bids for acquiring items of type $j$.  Then,

  \begin{enumerate}
    \item{\label{item:optimal_bids} (Optimal Bids for $j \in [M]$): In a second price auction, the dual variables $\mu_j$ are exactly equal to the optimal bids $x_j = \mu_j$.}

    \item{\label{item:pseudo_bids} (Pseudo Bids for $i \in [N]$): The dual variables $\rho_i, \mu_j$ satisfy the equalities $\mu_j = \underset{i \in \B_j}{\text{max}}\ v_{ij}\rho_i$, and $\rho_i = \underset{j \in \A_i}{\text{min}}\ \bigl(\frac{\mu_j}{v_{ij}}\bigr)$.}

    \item{\label{item:primal_allocation} (Recovering $R$): Dual optimal variables of Problem~\eqref{eqn:dual_problem} (\ie~the double dual variables) $\Gamma_{ij}$ (say) associated to the inequality $v_{ij}\rho_i \le \mu_j$ constitute an optimal allocation matrix $R = \Gamma$ for Problem~\eqref{eqn:main_problem}}
      
    \item{\label{item:useful_items} (Useful Items): The dual slack variables induce a collection $\A_i^\star = \{j \in \A_i\ |\ \theta_{ij} = 0\}$ which is such that $j \notin \A_i^\star \implies R_{ij} = 0$.  As well, the contrapositive: $R_{ij} > 0 \implies j \in \A_i^\star$}

    \item{\label{item:fulfilled_contracts} (Fulfilled Contracts): The dual slack variables induce a collection $\B_j^\star = \{i \in \B_j\ |\ \theta_{ij} = 0\}$ which is such that $i \not\in \B_j^\star \implies R_{ij} = 0$.  As well, the contrapositive: $R_{ij} > 0 \implies i \in \B_j^\star$.}
  \end{enumerate}
\end{theoremEnd}
\begin{proofEnd}
 \textit{Proof.}\ \ref{item:optimal_bids} follows by inspecting the derivation of $\L(s^\star, \ldots)$ in Equation~\eqref{eqn:lagrangian_s}, the Lagrangian after minimizing over $s$.  The optimum value of $s_j$ is obtained by minimizing $\lambda_j \Lambda_j(s / \lambda_j) - \mu_j s$, which occurs at $\Lambda_j'(s / \lambda_j) = \mu_j$.  From the fact $s = \lambda_j W_j(x)$ for the bid required to obtain supply $s$, we obtain the optimum bid through $\Lambda_j' \circ W_j(x) = \mu_j$ which results in $x_j = g_j^{-1}(\mu_j)$ (see Lemma~\ref{prop:bid_mapping_func}), and this is nothing but $x_j = \mu_j$ in the second price case.

  Conclusion \ref{item:pseudo_bids} follows from the monotonicity in the objective of Problem~\eqref{eqn:dual_problem}.  Indeed, since $-\Lambda_j^\star(\mu_j)$ is monotone decreasing in $\mu_j$, and the goal of the program is to \textit{maximize} the objective, is is optimal that $\mu_j$ be as small as possible while still satisfying the constraints $\forall j \in [M], i \in \B_j:\ v_{ij} \rho_i \le \mu_j$, which is exactly $\mu_j = \underset{i \in \B_j}{\text{max}}\ v_{ij} \rho_i$.  Similar reasoning applies for $\rho_i$ since $C_i > 0$, resulting in the requirement that $\rho_i$ be as large as possible while still satisfying the constraints $\forall i \in [N], j \in \A_i:\ \rho_i \le \frac{\mu_j}{v_{ij}}$ which is exactly $\rho_i = \underset{j \in \A_i}{\text{min}}\ \frac{\mu_j}{v_{ij}}$.

  The conclusions \ref{item:useful_items} and~\ref{item:fulfilled_contracts} follow from complementary slackness.  Recall the dual variable $\theta_{ij} = \mu_j - v_{ij} \rho_i$ from Equation~\ref{eqn:lagrangian_s}, which was eliminated before the statement of Problem~\eqref{eqn:dual_problem}.  We have $\theta_{ij} > 0 \iff \mu_j > v_{ij} \rho_i \iff \rho_i < \frac{\mu_j}{v_{ij}}$ (\ie~if $i$ or $j$ do not attain the respective maxima or minima from Item~\ref{item:pseudo_bids}).  The variable $\theta_{ij}$ is associated with the constraint $R_{ij} \ge 0$ and hence by complementary slackness $\theta_{ij} R_{ij} = 0$ so that $\theta_{ij} > 0 \implies R_{ij} = 0$.

  Finally, we establish Item~\ref{item:primal_allocation} by deriving the double dual of Problem~\eqref{eqn:dual_problem}.  We have the Lagrangian with dual multipliers $\Gamma_{ij} \ge 0$:

  \begin{align*}
    \L(\rho, \mu, \Gamma)
    &= \sum_{i = 1}^N \rho_i C_i - \sum_{j = 1}^M \lambda_j \Lambda_j^\star(\mu_j) + \sum_{i = 1}^N \sum_{j \in \A_i} \Gamma_{ij} (\mu_j - v_{ij} \rho_i)\\
    &= \sum_{i = 1}^N \rho_i \bigl(C_i - \sum_{j \in \A_i} \Gamma_{ij} v_{ij}\bigr) + \sum_{j = 1}^M\bigl[\mu_j\sum_{i \in \B_j} \Gamma_{ij} - \lambda_j \Lambda_j^\star(\mu_j) \bigr]\\
    &= \sum_{i = 1}^N \rho_i \bigl(C_i - \sum_{j \in \A_i} \Gamma_{ij} v_{ij}\bigr) + \sum_{j = 1}^M \lambda_j \bigl[\frac{\mu_j}{\lambda_j}\sum_{i \in \B_j} \Gamma_{ij} - \Lambda_j^\star(\mu_j) \bigr]\\
  \end{align*}
  Taking the supremum over $\rho$ results in the constraint $\sum_{j \in \A_i} \Gamma_{ij} v_{ij} = C_i$ and subsequently taking the supremum over $\mu_j$ results in

  \begin{align*}
    \underset{\rho, \mu}{\text{sup}}\ \L(\rho, \mu, \Gamma) =
    \begin{cases}
      \sum_{j = 1}^M \lambda_j \Lambda_j^\star\bigl(\frac{1}{\lambda_j}\sum_{i \in \B_j}\Gamma_{ij}\bigr) & \text{if } \sum_{j \in \A_i}\Gamma_{ij} v_{ij} = C_i\\
      \infty & \text{otherwise,}
    \end{cases}
  \end{align*}
  since ${\bigl(\Lambda_j^\star\bigr)}^\star = \Lambda_j$ by the Fenchel-Moreau Theorem (see \eg~\cite[Prop 1.6.1]{bertsekas2009convex} or~\cite[Ch. E]{hiriart2012fundamentals}).  Thus, if we add the linear equality constraint $s_j = \sum_{i \in \B_j} \Gamma_{ij}$ we are left with exactly the primal Problem~\eqref{eqn:main_problem}.  Thus, the optimal dual variables $\Gamma_{ij}$ constitute an optimal allocation array $R_{ij}$ to the primal problem.
\end{proofEnd}

Let us consider a number of corollaries to this result.  We will discuss interpretations of these corollaries of Proposition~\ref{prop:duality_consequences} further in Section~\ref{sec:interpretations_of_duality}.

\begin{theoremEnd}[normal]{corollary}[Induced Valuations]%
  \label{cor:induced_valuations}
  For each fixed $i \in [N]$, for any item type $j \in \A_i^\star$ the optimal bid is a positive multiple of the pseudo-bid $\rho_i$: $\mu_j = v_{ij} \rho_i$.
\end{theoremEnd}

If item valuations take values only $1$ or $0$, then there is just a single optimum bid across each of the sets $\A_i^\star$.

\begin{theoremEnd}[normal]{corollary}[Uniform Bids]%
  \label{cor:split_graph_ubp}
Suppose that $v_{ij} \in \{0, 1\}$.  Then, the optimal bids placed in a second price auction across all items of types $j \in \A_i^\star$ are equal and given by $x_j = \rho_i$.
\end{theoremEnd}
\begin{proofEnd}
 \textit{Proof.}\ The dual variables $\mu_j$ are equal to the optimal bids for items of type $j$: $x_j = \mu_j$.  Then, by the definition of $\A_i^\star$ we must have $\theta_{ij} = \mu_j - \rho_i = 0$.
\end{proofEnd}

In the particularly simple case where items do not have differing valuations, and any item can be used to fulfill any contract, then we can recognize that there is just a single optimal pseudo bid characterizing all of the bids, as well as that, for second price auctions, there is just one single optimal bid.

\begin{theoremEnd}[normal]{corollary}[Uniform Bid]%
  \label{cor:ubp}
  If $v_{ij} = 1$ for each $i, j$, then there exists a single pseudo bid $\rho^\star$ such that the optimal bids satisfy $x_j = \rho^\star$ across all $j \in [M]$.
\end{theoremEnd}
\begin{proofEnd}
  \textit{Proof.}\ We must have $\rho_i = \text{min}_{j \in [M]} \mu_j$, and since the r.h.s.\ does not depend on $i$, there must be some $\rho^\star$ such that $\rho_i = \rho^\star$ for each $i$.
\end{proofEnd}

\begin{remark}
  In the special case where Corollary~\eqref{cor:ubp} holds, the problem reduces into a monotone function root-finding problem for $\rho \mapsto \sum_{j = 1}^M \lambda_j W_j (\rho) - \sum_{i = 1}^N C_i$.  The structure of the solution when there is only one contract $N = 1$ is comparable.
\end{remark}

\subsubsection{Interpretations}%
\label{sec:interpretations_of_duality}

  \paragraph{Induced Valuations and Supply and Demand}
  In many game theoretic analyses of auction markets, a \textit{valuation} over items is assumed to exist for market participants, and this valuation is essential to the analysis, since it is incorporated into the agent's utility functions.  While our problem formulation contains item valuations $v_{ij}$, these are not valuations in the same game theoretic sense.  Indeed, the intermediary does not care at all about the items themselves, only about fulfilling their contractual obligations.  However, in second price auctions with private valuations, it is a dominant strategy simply to \textit{bid your valuation}~\cite{krishna2009auction}.  Thus, Corollary~\ref{cor:induced_valuations} can be understood as a way in which item valuations are \textit{induced} by the contractual requirements.  For example, if we have $v_{ij} \in [0, 1]$ and we interpret these quantities as conversion probabilities (in advertising, a conversion is a decision to make a purchase) then the equation $\mu_j = v_{ij} \rho_i$, with $\mu_j$ being the optimal bid for items of type $j$, indicates that $\rho_i$ is the value of a single conversion of type $j$ for contract $i$.

Moreover, if we recall from the Lagrangian that the dual variables $\rho_i$ are associated to the constraint $\sum_{j = 1}^M \lambda_j R_{ij} v_{ij} = C_i$, the \textit{shadow price interpretation} tells us that $\rho_i$ is equal to the marginal cost to obtaining additional conversions for contract $i$, and can thus be used by the DSP as part of the process of determining what price they should sell contracts for.

Similarly, the conclusion~\ref{item:optimal_bids} drawn from Proposition~\ref{prop:duality_consequences}, that the dual multiplier $\mu_j$ associated with the constraint $\sum_{i = 1}^N R_{ij} = s_j$ is exactly the optimal bid $\mu_j = x_j$ can be understood, as in the proof, by examining the analysis of the Lagrangian function.  In particular, there arises the minimization over $s_j$ of the function $s_j \mapsto \lambda_j \Lambda_j(s_j / \lambda_j) - \mu_j s_j$, the solution of which occurs when $\Lambda_j'(s_j / \lambda_j) = \mu_j$ and since $\Lambda_j' = W_j^{-1}$ we must have $s_j = \lambda_j W_j(\mu_j)$.  Indeed, these relations indicate that $\mu_j$ is exactly equal to the marginal cost of obtaining items of type $j$ at the acquisition rate $s_j$.  And, in the case of a second price auction, this is what one must bid in order to win the item against the marginal competing bidder.

To summarize some of this discussion, one may keep in mind that, for second price auctions and with $v_{ij} \in [0, 1]$ being conversion probabilities, we can interpret $\mu_j$ as being the value to the intermediary of obtaining an impression of type $j$, and $\rho_i$ as the value to contract $i$ of obtaining a conversion.  These valuations are \textit{induced} through an optimal balance between the supply available (as characterized by $\lambda_j, W_j$) and the supply demanded (as characterized through $v_{ij}, C_i$).

\paragraph{Graph Partitioning}
Continuing the discussion to look at conclusion~\ref{item:useful_items} and~\ref{item:fulfilled_contracts} of Proposition~\ref{prop:duality_consequences}, these  establish existence of subsets $\A_i^\star \subseteq \A_i$ and $\B_j^\star \subseteq \B_j$ constituting, respectively, the items which are \textit{actually used} for the fulfillment of contract $i$, and the contracts towards which items of type $j$ are \textit{actually sent}.  These sets are a result of complementary slackness applied to the quantity $\theta_{ij} = \mu_j - v_{ij}\rho_i$.  If $\theta_{ij} > 0$ it indicates that the cost for contract $i$ to bid on items of type $j$ is greater (by the amount $\theta_{ij}$) than the minimum cost at which conversions can be acquired for that contract, and hence type $j$ should not be used to fulfill contract $i$: $R_{ij} = 0$.

These sets $\A_i^\star, \B_j^\star$ (more directly, the slack variables $\theta_{ij}$) induce partitions of the graph $\mathcal{G} = \{(i, j)\ |\ i \in [N], j \in \A_j\}$ as in $\mathcal{G}_i^\star \defeq \{(i, j)\ |\ i \in [N], j \in \A_i^\star\}$ being the subgraph of $\mathcal{G}$ to which contract $i$ belongs.  Within this subgraph, bids are determined entirely by the scalar $\rho_i$ and the values ${\bigl(v_{ij}\bigr)}_{j \in \A_i^\star}$.  In fact, due to Corollary~\ref{cor:split_graph_ubp} if $v_{ij} \in \{0, 1\}$ the bids placed on items in this subgraph are all equal to $\rho_i$.  This graph partitioning was also recognized in~\cite{marbach_bidding_2020}, but not as a consequence of convex duality.

Finally, in Corollary~\ref{cor:ubp}, we recognize that if there is no differentiation between item valuations, and any item can be used to fulfill any contract, \ie~$v_{ij} = 1$, then there is no reason to place different bids for different items.  The existence of such a $\rho^\star$ holds in general only in this very special case.  Indeed, if $v_{ij} \in \{0, 1\}$, the availability to contract $i$ of item types with lower costs can induce differing pseudo bids across contracts.  However, we have observed in some computational examples that, particularly when there is little margin in Assumption~\ref{ass:adequate_supply} (\ie~there is just barely enough supply available to fulfill the contracts), the optimal solution may still collapse to the case $\forall i \in [N]\ \rho_i = \rho^\star$ (an illustration of this phenomena is given in Section~\ref{sec:computational_examples}).

\begin{remark}[Sets $\A_i^\star, \B_j^\star$ in Practice]%
  \label{rem:AB_tolerance}
  When obtaining solutions to Problem~\eqref{eqn:dual_problem} numerically, it may be the case only that $\theta_{ij} \approx 0$, up to the tolerance parameters set for the algorithm.  In this case, and considering the interpretation provided in the previous paragraph, it may be reasonable to fix some parameter $\epsilon > 0$ and treat the slack variables $\theta_{ij}$ as if they are equal to zero whenever they satisfy $\theta_{ij} \le \epsilon$.  Then, the intermediary would be overpaying by at most $\epsilon$.
\end{remark}

\subsection{Computational Examples}%
\label{sec:computational_examples}
In this section we examine some computational simulation examples.  The primary purpose of this section is to further develop an intuitive and qualitative understand of the contract management problem.  Furthermore, these examples demostrate the algorithmic implications of convexity --- there exists efficient algorithms and software that can solve non-trivial instances of the contract management problem.  In Section~\ref{sec:bidding_bifurcation} we illustrate how the conclusion of Corollary~\ref{cor:ubp} (the existence of a single optimal bid $\rho^\star$) may sometimes hold for $v_{ij} \in \{0, 1\}$, and in Section~\ref{sec:large_scale_example} we illustrate a large scale example with over $1000$ item types and the sparsity resulting through $\theta_{ij}$.  All of our simulations are carried out using Python's scientific computing stack~\cite{2020SciPy-NMeth} and \texttt{cvxpy}~\cite{diamond2016cvxpy}.  It is important to recognize that Problem~\eqref{eqn:main_problem} is not always an instance of a particular well known class of convex optimization problems (\eg~linear program, quadratic program, \etc), particularly if $W_j$ is estimated through kernel smoothing or other generic method.  Thus, the derivation of specialized algorithms is important for practical application; this is a topic of ongoing work.

\subsubsection{Bidding Bifurcations}%
\label{sec:bidding_bifurcation}

\begin{figure}
  \centering

  \begin{subfigure}[b]{0.4\textwidth}
    \includegraphics[width=\textwidth]{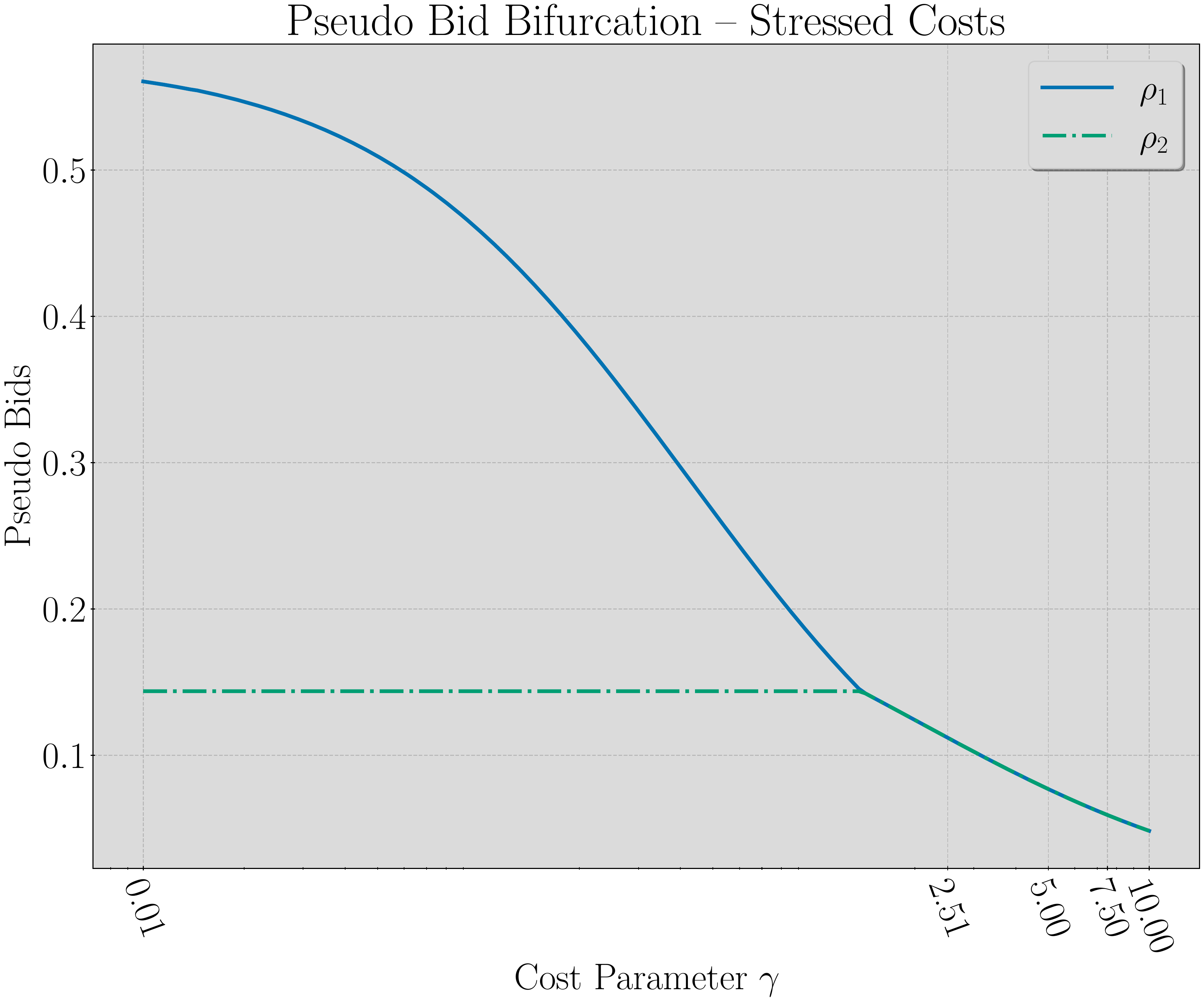}
    \caption{Stressing Cost Differences}%
    \label{fig:bidding_bifurcations_cost}
  \end{subfigure}
  \begin{subfigure}[b]{0.4\textwidth}
    \includegraphics[width=\textwidth]{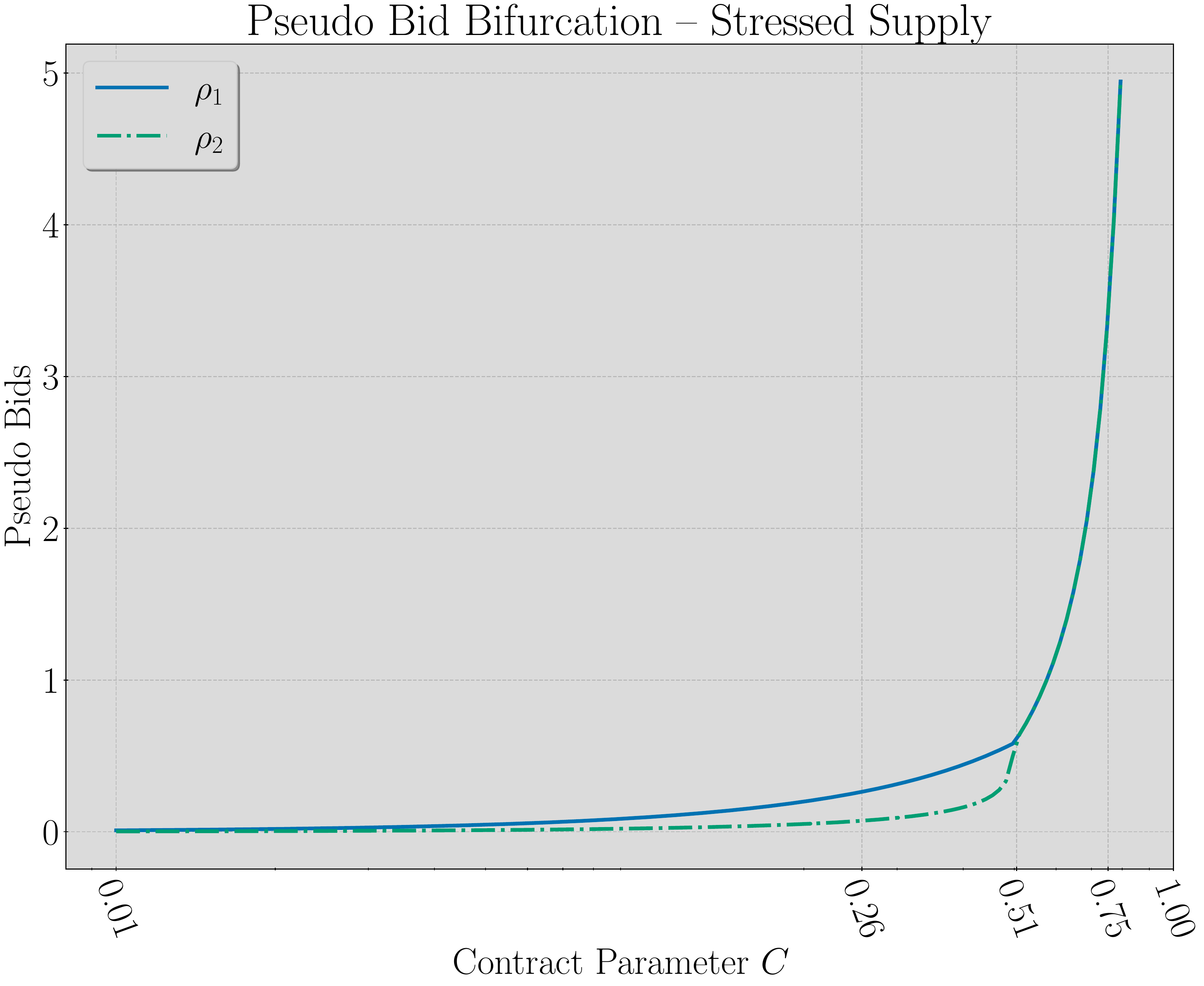}
    \caption{Stressing Available Supply}%
    \label{fig:bidding_bifurcations_supply}
  \end{subfigure}

  {\small Examples of how optimal bids can naturally differ between contracts and items, as well as examples of the cases where the Uniform Bid Principle (Corollary~\ref{cor:ubp}) tends to hold.}
  \caption{Bifurcation Examples of Pseudo Bids}%
  \label{fig:bidding_bifurcations}
\end{figure}

Even though Corollary~\ref{cor:ubp} holds only in very special circumstances, it turns out that there is still just a single optimal bid applying across all item types in some practical scenarios as well.  Typically this arises when the available rate of items is only just adequate the fulfill contracts (\ie~there is little margin available in Assumption~\ref{ass:adequate_supply}), but we also observe that it can occur when contracts have shared access to a cheap source of items.  These cases are exemplified in Figure~\ref{fig:bidding_bifurcations}.

To construct Figure~\ref{fig:bidding_bifurcations} we have considered a simple case when $M = 3$ item types and $N = 2$ contracts.  In both subfigures we have $v_{ij} \in \{0, 1\}$ with $\A_1 = \{1, 2\}$, and $\A_2 = \{2, 3\}$ and $\lambda_1 = \lambda_2 = \lambda_3 = 1$.  As well, the supply curves are given by exponentials $W_j(x) = 1 - e^{-\gamma_j x}$ (see Figure~\ref{fig:example_functions}).  The average cost of these items is given by $1 / \gamma_j$, and hence become cheaper as $\gamma$ becomes larger.

In Figure~\ref{fig:bidding_bifurcations_cost} we have $\gamma_1 = 1/2, \gamma_3 = 2$ but $\gamma_2 = \gamma$ is a parameter.  We see that, as item type $2$ gets cheaper (\ie~$\gamma$ increases), it is eventually the case that both contracts draw the majority of their items from type $2$.  Prior to this, the majority of items are drawn from the cheaper types $1$ for contract $1$ or $3$ for contract $2$.  It should be pointed out that if $\gamma_1 = \gamma_3$, the contracts would still draw the majority of their items from types $1$ and $3$, respectively, for small values of $\gamma_2$, but their bids would still be identical as a coincidence of having access to different items that happen to have the same costs.  That is, the reason that bids coincide in that case would be a simple pathology of particular numerical values, not a consistent structural feature.

In Figure~\ref{fig:bidding_bifurcations_supply} we have $\gamma_1 = 1/10$ (expensive), $\gamma_2 = 1$, and $\gamma_3 = 10$ (cheap).  As well, we modify the contract item requirements $C_1 = C$ and $C_2 = 2C$.  We see that, as $C$ approaches $1$, and the available supply is stressed, the bids are eventually equal (and large), even though there are large differences between the costs of items available to the contracts.

\subsubsection{Large Scale Example and Dual Induced Sparsity}%
\label{sec:large_scale_example}

In this section we provide a large scale randomized example.  This demonstrates the potential for scaling the size of Problem~\eqref{eqn:main_problem} to large and practical instances (\eg~thousands of contracts and item types), as well as to provide qualitative illustrations of the sparsification in induced item valuations through the slack variable $\theta$.

\begin{figure}
  \centering

  \begin{subfigure}[b]{0.4\textwidth}
    \includegraphics[width=\textwidth]{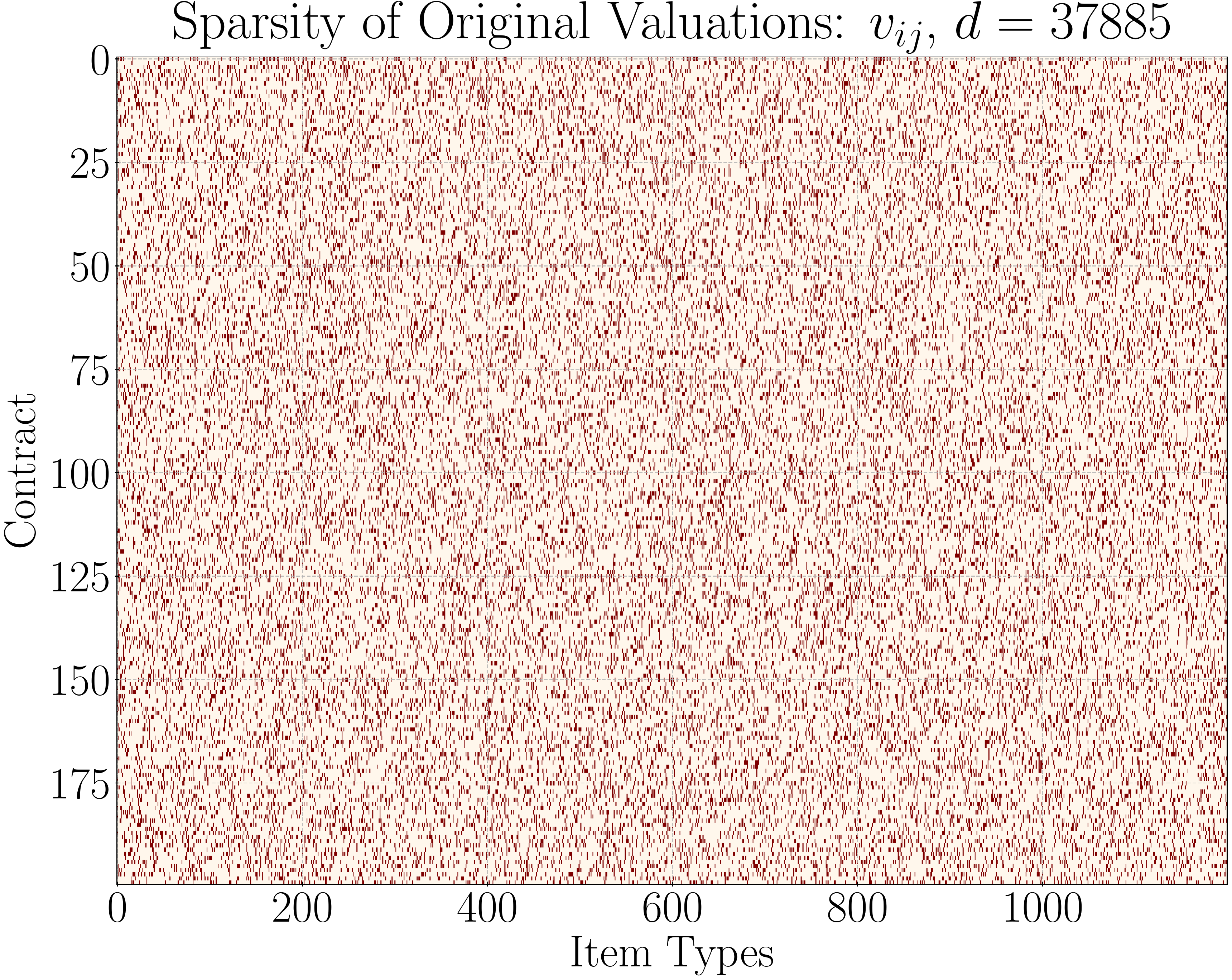}
    \caption{Original Valuations}%
    \label{fig:original_valuations}
  \end{subfigure}
  \begin{subfigure}[b]{0.4\textwidth}
    \includegraphics[width=\textwidth]{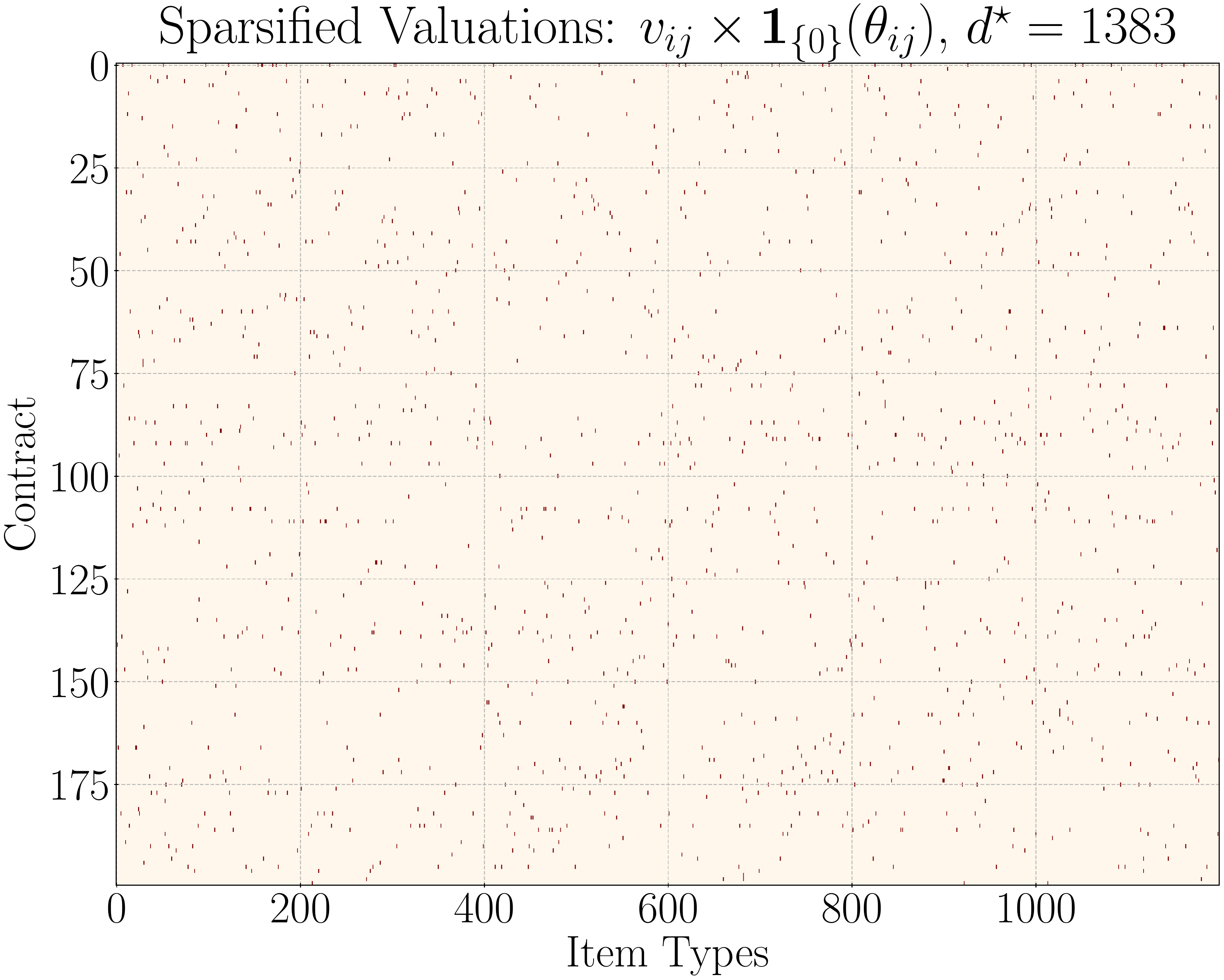}
    \caption{Sparsified Valuations}%
    \label{fig:sparsified_valuations}
  \end{subfigure}
  \caption{Dual Induced Sparsification}%
  \label{fig:sparsification_example}

  {\small A large scale example with $M = 1200$, $N = 200$.  The parameters $\lambda_j$, $\gamma_j$ (for $W_j(x) = 1 - e^{-\gamma_j x}$) and $C_i$ are all $\text{exp (1)}$ distributed.  As well, we construct valuations via $v_{ij} \sim \mathcal{N}(0, 1) - 1$ and then clipping the values between $[0, \infty)$.}
\end{figure}

Moreover, when $\theta_{ij} > 0$, the item $j$ is not used for contract $i$, which induces the subsets $\A_i^\star \subseteq \A_i$.  To get a sense of this effect, consider Figure~\ref{fig:sparsification_example} where we display the sparsity pattern of $v_{ij}$ as well as $v_{ij} \times \ind_{\{0\}}(\theta_{ij})$.  Additionally, the figure points out the difference between $d = \sum_{i = 1}^N |\A_i|$ and $d^\star = \sum_{i = 1}^N |\A_i^\star|$.  The former is the \textit{basic} sparsity of the problem (which is already $d \ll MN$, in this case $MN = 240,000$) and $d^\star$ is the \textit{induced} sparsity of the problem.  While it naturally depends on the particulars of $v, \lambda, W$ etc., it can be the case that $d^\star \ll d$, and hence the dual program can induce substantial additional sparsity into the matrix $R$.  Of course, these specific results are highly dependent upon the particulars of the simulation setup, but the examples serve to illustrate the predictions made by Proposition~\ref{prop:duality_consequences}.

\section{The First Price Case}
\label{sec:first_price_case}
The cost of bidding in first price auctions is easy to define:

\begin{equation}
\label{eqn:first_price_cost}
  f_{1st}(x) = \E[x \ind(p \le x)] = xW(x),
\end{equation}
since the bidder pays their bid $x$ when they win, and they have a probability $W(x)$ of winning.  The acquisition cost can also easily be recognized

\begin{equation}
  \label{eqn:acquisition_cost_auction_1pa}
  \begin{aligned}
    \Lambda_{1st}(q) &= q W^{-1}(q).
  \end{aligned}
\end{equation}
Unlike the second price case, it is possible that that $\Lambda(q) \rightarrow \infty$ as $q \rightarrow 1$.  This occurs if the maximum bid $\bar{x} = \infty$.

Like the second price case, we are interested in the convexity of the function $\Lambda_{1st}$.  In first price auctions however, this does not hold unconditionally -- some additional assumptions are needed to establish the convexity of $\Lambda_{1st}$.  We first need the following definition.

\begin{definition}[$\alpha$-concavity]%
  \label{def:alpha_concavity}
  Define, for $\alpha \ge 0$, $x > 0$ the function

  \begin{equation*}
    \ell_\alpha(x) \defeq \int_1^x \frac{1}{t^\alpha}\d t =
    \left\{\begin{array}{lr}
             \ln x & \alpha = 1\\
             \frac{x^{1 - \alpha} - 1}{1 - \alpha} & \text{otherwise}
           \end{array}\right.,
  \end{equation*}
  where in particular $\ell_2(x) = 1 - 1/x$.  We will say that a
  positive function $W: \R \rightarrow (0, \infty)$ is (strictly)
  $\alpha$-concave if $\ell_\alpha \circ W$ is (strictly)
  concave.  In particular, $W$ is log-concave if $\alpha = 1$ and
  concave if $\alpha = 0$.
\end{definition}

It is shown in the appendix, Proposition~\ref{prop:alpha_log_concavity}, that the condition required to establish convexity in the first price case, namely $2$-concavity, is a \textit{weaker} condition than is log-concavity.  That is, the log-concavity of $W$ is a sufficient (but not necessary) condition for convexity in the first price case; it is well known that most famous distribution functions are log-concave.

%% HIERARCHY LEMMA HERE MOVED TO APPENDIX MANUALLY

\begin{theoremEnd}[normal]{proposition}[Convex Acquisition Costs --- First Price Case]%
  \label{prop:convex_acquisition_costs_1pa}
  Suppose that the supply curve $W(x)$ is strictly $2$-concave (\cf~Definition~\ref{def:alpha_concavity}), \ie~$\ell_2 \circ W$ is strictly concave on $(0, \infty)$ where $\ell_2(x) = 1 - 1/x$.  Then in a first price auction, the extended acquisition cost function

  \begin{equation}
    \label{eqn:acq_costs_1pa}
    \Lambda_{1st}(q) \defeq \left\{\begin{array}{lr}
                               \infty; & q > 1\\
                               0; & q < 0\\
                               qW^{-1}(q); & \mathrm{otherwise}
                             \end{array}\right.,
  \end{equation}
  is a proper, lower semi-continuous, non-decreasing, and convex function on $\R$.  Moreover, $\Lambda_{1st}$ is strictly convex over $[0, 1]$.
\end{theoremEnd}
\begin{proofEnd}
  Since $f(x) = xW(x)$ we have, for $q \in (0, 1)$, $\Lambda(q) = q W^{-1}(q)$.  On $q \le 0$ we define $\Lambda(q) = 0$, on $q > 1$ we define $\Lambda(q) = \infty$, and finally $\Lambda(1) = \bar{x} \le \infty$.

Convexity will therefore follow if $\Lambda$ is convex on $(0, 1)$.  To this end, we use the $2$-concavity of $W$ to see that $1 - 1 / W(x)$ is concave on its domain and therefore that the inverse, $W^{-1}\bigl(1 / (1 - x)\bigr)$ is convex on $0 < \frac{1}{1 - x} < 1$.  It is well known that for a convex function $g$, the function

  \begin{equation*}
    (cx + d) g\bigl(\frac{ax + b}{cx + d}\bigr)
  \end{equation*}
  is convex on $cx + d > 0$ (see e.g.~\cite[Ex. 3.20]{boyd2004convex}). Therefore, by setting $a = c = 1$, $b = -1$ and $d = 0$ we obtain convexity of

  \begin{align*}
    (cq + d) W^{-1}\Bigl(\frac{1}{1 - \frac{aq + b}{cq + d}}\Bigr)
    &= qW^{-1}\bigl(\frac{1}{1 - \frac{q - 1}{q}}\bigr)\\
    &= qW^{-1}\bigl(q\bigr)
  \end{align*}
  which is the function $\Lambda(s)$.  Since $qW^{-1}(q)$ is strictly monotone increasing on $(0, 1)$, $\Lambda_{1st}$ is strictly convex on this interval, and evidently non-decreasing on all of $\R$.  
\end{proofEnd}

In the first price case, it is extremely useful to assume some additional regularity of $W$, namely that it is differentiable.

\begin{definition}[Differentiable Supply Curve]%
  A differentiable supply curve $W$ is a supply curve (\cf~Definition \ref{def:supply_curve}) which is continuously differentiable on $(0, \bar{x})$ with derivative $W'$.
\end{definition}

Additionally, the first price case gives rise to what we refer to as the \textit{bid-mapping function}.  This is a function $g$ that is is \textit{defined} such that $\Lambda'_{1st}(q) = g \circ W^{-1}(q)$.  That is, $g$ is such that the inverse of the derivative of the acquisition cost curve is given through the composition of the supply curve with the function $g^{-1}$.  In a second price auction, since $\Lambda'_{2nd}(q) = W^{-1}(q)$, and therefore $\Lambda'^{-1}(x) = W(x)$, the bid mapping function is simply the identity.  Thus, the function $g$ is relevant only for first price auctions.  We can write an expression for the bid mapping function in first price auctions.

\begin{theoremEnd}[normal]{proposition}[Bid Mapping Function -- First Price Auctions)]%
  \label{prop:bid_mapping_func}
  Let $W$ be a differentiable supply curve.  The bid mapping function $g$, such that $\Lambda'_{1st}(q) = g \circ W^{-1}(q)$, is given by

  \begin{equation*}
    g(x) =
    \begin{cases}
      x + \frac{W(x)}{W'(x)}, & \text{if } x \in (0, \bar{x})\\
      \bar{x} + {\bigl(\underset{x \rightarrow \bar{x}}{\lim}\ W'(x)\bigr)}^{-1}, & \text{if } x = \bar{x}\\
      0, & \text{if } x \le 0\\
      \infty, & \text{if } x > \bar{x}.
    \end{cases}
  \end{equation*}
  Moreover, if $W$ is strictly $2$-concave (See Proposition~\ref{prop:convex_acquisition_costs_1pa}) then $g$ is a strictly monotone increasing function on $[0, \bar{x}]$ with range $[0, g(\bar{x})]$, where $g(\bar{x}) = \infty$ if $\bar{x} = \infty$.  Finally, $g(x)$ is a continuous function.
\end{theoremEnd}
\begin{proofEnd}
  We have the calculations

  \begin{align*}
    \frac{\d}{\d q} \Lambda(q)
    &= \frac{\d}{\d q} q W^{-1}(q)\\
    &= W^{-1}(q) + \frac{q}{W' \circ W^{-1}(q)}.
  \end{align*}
  Therefore, $\Lambda_{1st}'(q) = g_{1st} \circ W^{-1}(q)$ where $g_{1st}(x) = x + \frac{W(x)}{W'(x)}$.  Moreover, when $\Lambda_{1st}(q)$ is convex (see Proposition~\ref{prop:convex_acquisition_costs_1pa}) the derivative $\Lambda_{1st}(q)$ is monotone and therefore, since $g_{1st} = \Lambda_{1st}' \circ W$, it is monotone, since the composition of monotone increasing functions is monotone increasing.  We fill in the value of $g$ at $\bar{x}$ by continuity, and $g$ is continuous at $x = 0$ since $\frac{d}{\d x} \ln W(x) = W'(x) / W(x)$ and $\ln W(x) \rightarrow -\infty$ as $x \rightarrow 0$, so it must be that the derivative of $\ln W(x)$ is converging to $\infty$ and therefore $W(x) / W'(x) \rightarrow 0$ as $x \rightarrow 0$.  
\end{proofEnd}

The bid mapping function facilitates an expression for the conjugate of $\Lambda_{1st}$.
  
\begin{theoremEnd}[normal]{proposition}[Fenchel Conjugate --- First Price Case]%
  \label{prop:acquisition_cost_duality_1pa}
  Let $\Lambda(q) = f_{1st} \circ W^{-1}(q)$ where $W$ is a strictly $2$-concave differentiable supply curve.  The Fenchel conjugate $\Lambda^\star(\mu)$ is given by

  \begin{equation}
    \Lambda^\star(\mu) =
    \begin{cases}
      \infty, & \text{if } \mu < 0\\
      \bigl(\mu - g^{-1}(\mu)\bigr) W \circ g^{-1}(\mu), & \text{if } \mu \in [0, g(\bar{x})]\\
      \mu - \bar{x}, & \text{if } \mu > g(\bar{x}).
    \end{cases}
  \end{equation}
  The function $\Lambda^\star$ is a proper, convex, and lower-semicontinuous function, which is strictly convex and strictly monotone increasing on $\R_+$.
\end{theoremEnd}
\begin{proofEnd}
  By definition, we need to calculate \[\Lambda^\star(\mu) = \underset{q \in (-\infty, 1]}{\text{sup}}\ [\mu q - \Lambda(q)].\]  For $\mu \in [0, g(\bar{x})]$ we apply Fermat's rule to see that we need to solve $\mu = \Lambda'(q)$ for $q$ and whence we obtain $q = W \circ g^{-1}(\mu)$, where the inverse $g^{-1}$ exists by Proposition~\ref{prop:bid_mapping_func}.  Substituting this into the definition we have

  \begin{align*}
    \mu q - \Lambda(q)
    &= \mu W \circ g^{-1}(\mu) - f_{1st} \circ W^{-1} \circ W \circ g^{-1}(\mu)\\
    &= \mu W \circ g^{-1}(\mu) - g^{-1}(\mu) W \circ g^{-1}(\mu)\\
    &= \bigl(\mu - g^{-1}(\mu)\bigr) W \circ g^{-1}(\mu).
  \end{align*}
  Finally, if $\mu > g(\bar{x})$ there is no solution to this system and we have $q = 1$ by monotonicity.

  That $\Lambda^\star$ is strictly convex on $\R_+$ follows since $\Lambda_{1st}$ is differentiable on $(0, 1)$ and Theorem~\cite[11.13]{rockafellar1998variational} establishing a relationship between differentiability and strict convexity of convex functions and their conjugates.

  The strict monotonicity of $\Lambda^\star$ on $\mu \ge \bar{x}$ is clear, so it remains to show that it is strictly monotone on $[0, \bar{x})$.  To this end, recognize that $\Lambda^\star \circ g(x) = \bigl(g(x) - x\bigr) W(x)$ and this is equal to $W(x)^2 / W'(x)$ on the interval $[0, \bar{x})$ by the definition of $g(x)$.  Then, since $W$ is strictly $2-concave$ it must be that the derivative of $1 - 1 / W(x)$ is strictly decreasing, and thus $W'(x) / W(x)^2$ is a decreasing function.  Therefore, $\Lambda^\star \circ g(x)$ is strictly increasing $[0, \bar{x})$.  Since $g$ is strictly increasing, and the composition $\Lambda^\star \circ g$ is strictly increasing, it follows that $\Lambda^\star$ is as well.  
\end{proofEnd}

\paragraph{Contract Management in First Price Auctions}
In light of Proposition \ref{prop:convex_acquisition_costs_1pa}, we can recognize that most of the analysis of Section \ref{sec:contract_management} continues to hold, as long as $W_j$ is a $2$-concave supply curve.  That is, Proposition \ref{prop:convex_reformulation} holds under this additional assumption.  As well, the consequences summarized in Proposition \ref{prop:duality_consequences} continue to hold as well, but for a single modification: the optimal bid $x_j$ is no longer equal to the dual multiplier $\mu_j$, rather, it must be obtained through the bid-mapping function as in $x_j = g_j^{-1}(\mu_j)$.  We formally summarize this in a proposition.

\begin{theoremEnd}[normal]{proposition}[Contract Management in First Price Auctions]
  In a first price auction, let $W_j$ be a strictly $2$-concave differentiable supply curve and suppose that Assumption~\ref{ass:adequate_supply} holds.  Then, Problem \eqref{eqn:main_problem_Wf} can be reformulated as a convex program \eqref{eqn:main_problem} and the conclusions of Proposition \ref{prop:convex_reformulation}, and the existence and regularity of solutions Proposition~\ref{prop:regularity}, continue to hold.

  Moreover, the consequences of duality, Proposition~\ref{prop:duality_consequences} continue to hold verbatim, except Item~\ref{item:optimal_bids}, which must be modified to say that if $W_j$ is additionally a differentiable supply curve, then the optimal bids $x_j$ are obtained through the bid-mapping function $x_j = g_j^{-1}(\mu_j)$.
\end{theoremEnd}
\begin{proofEnd}
  The proof of Proposition \ref{prop:convex_reformulation} depends only upon the convexity of $\Lambda_j$, which holds for first price auctions when $W_j$ is $2$-concave by Proposition~\ref{prop:convex_acquisition_costs_1pa}.  Similarly, the analysis leading up to Problem~\eqref{eqn:dual_problem} does not depend upon the auction type.  The consequences, other than Item~\ref{item:optimal_bids}, described in Proposition~\ref{prop:duality_consequences} follow from strict the monotonicity of $\Lambda_j^\star$, which holds under the strict $2$-concavity of $W_j$ by Proposition \ref{prop:acquisition_cost_duality_1pa}.

To modify Item~\ref{item:optimal_bids} of that proposition, we again inspect the derivation of $\text{inf}_s\ \L$ in Equation \eqref{eqn:lagrangian_s}.  When $W_j$ is differentiable, the optimum bid is obtained through the solution of $\Lambda_j' \circ W_j(x) = \mu_j$, which occurs at $x = g_j^{-1}(\mu_j)$ in the first price case, by the definition of $g_j$.  
\end{proofEnd}

\section{Related Problems}
\label{sec:related_problems}
\subsection{Budget Constrained Optimal Bidding}%
The papers~\cite{zhang2014optimal} and~\cite{balseiro2015repeated} are important papers in RTB.  We will analyze closely analogous versions of their problems, modified just enough to maintain consistent notation and terminology with the rest of the present paper.  In each case, the objective is to calculate optimal bids $x \in \R^M$ so as to maximize the expected value of items obtained $\sum_{j = 1}^M \lambda_j v_j W_j(x_j)$ subject to a budget constraint: $\sum_{j = 1}^M \lambda_j f_j(x_j) \le B$ for some $B > 0$.  In~\cite{zhang2014optimal} it is a first price auction $f_j(x) = xW_j(x)$ and in~\cite{balseiro2015repeated} it is second price $f_j(x) = \int_0^x u\d W_j(u)$.  That is, we are faced with the problem

\begin{equation}
  \label{eqn:budget_constrained_bidding}
  \begin{aligned}
    \underset{x}{\text{minimize}} &\quad -\sum_{j = 1}^M \lambda_j v_j W_j(x_j)\\
    \text{subject to} &\quad \sum_{j = 1}^M \lambda_j f_j(x_j) \le B.
  \end{aligned}
\end{equation}
By making the invertible substitution $s_j = \lambda_j W_j(x_j)$ we can transform this into a convex optimization problem involving the acquisition cost functions $\Lambda_j$.  We have:

\begin{equation}
  \label{eqn:convex_budget_constrained_bidding}
  \begin{aligned}
    \underset{s}{\text{minimize}} &\quad -\sum_{j = 1}^M v_j s_j\\
    \text{subject to} &\quad \sum_{j = 1}^M \lambda_j \Lambda_j(s_j / \lambda_j) \le B.
  \end{aligned}
\end{equation}
One of the key results of~\cite{balseiro2015repeated} is to prove strong duality for Problem~\eqref{eqn:budget_constrained_bidding}, which seemed to be a surprising example where strong duality is obtained for a non-convex problem.  However, having recognized the hidden convexity in this problem with~\eqref{eqn:convex_budget_constrained_bidding}, strong duality is to be expected.

To calculate the optimal bids, consider the Lagrangian function \[\L(s, \vartheta) = -\vartheta B + \sum_{j = 1}^M \bigl[\vartheta \Lambda_j(s_j / \lambda_j) - v_j s_j \bigr]\] and apply the first order stationarity conditions to obtain the equation $\Lambda_j'(s_j / \lambda_j) = v_j / \vartheta$, which characterizes the optimum bid by $x_j^\star = g_j^{-1}(v_j / \vartheta)$.  This bid is exactly the shaded valuation $x_j^\star = v_j / \vartheta$ for second price auctions (since $g$ is the identity in that case), a typical result for budget constrained bidding~\cite{jiang2014bidding, gummadi2013optimal, balseiro2015repeated}.  In the first price case, as long as $W_j$ is strictly $2$-concave, the function $g_j^{-1}$ is strictly monotone increasing, and hence $x_j^\star$ is well defined.  To calculate the optimal Lagrange multiplier $\vartheta \ge 0$, we need only solve the monotone root-finding problem $\sum_{j = 1}^M \lambda_j f_j \circ g_j^{-1}(v_j / \vartheta) = 0$ for $\vartheta$, which can be accomplished through bisection.

The essence of the results of~\cite{zhang2014optimal} can be reproduced for the function $W_j(x) = \frac{x}{c_j + x}$ by carrying out the calculations $g_j(x) = x + W_j(x) / W_j'(x)$ and then obtaining $g_j^{-1}(v_j / \vartheta) = c_j\bigl(\sqrt{1 + \frac{v_j}{c_j\vartheta}} - 1\bigr)$.

\subsection{Limit Order Book Aware Markowitz Portfolio}
\label{sec:lob_problem}
Modern financial markets are organized around a \textit{limit order book} (LOB) that keeps track of the willingness of market participants to buy or sell securities at a certain price, and up to a certain volume~\cite{bouchaud2018trades}.  Because there is no single seller, the ``price'' of a security is ambiguous, but generally the quoted price is given by the \textit{mid price}, call it $m$.  That is, the average between the seller with the lowest willingness to sell (called the best ask), and the buyer with the greatest willingness to buy (called the best bid).  However, it is not in general possible to actually transact at this quoted mid price, since there is no willing counter-party (if there were, then it wouldn't be the mid price) and the amount actually paid to acquire (or the amount obtained by selling) $V$ shares must be strictly greater (lesser, for selling) than the nominal value $mV$.  This difference is the \textit{transaction cost}, and for concreteness, we will focus on the case of \textit{buying} --- the case of selling being exactly symmetric.

There are numerous sources of transaction costs --- the first is the spread (simply the difference between the mid price and the best bid), but there is also a \textit{volume cost} that arises when there are fewer than $V$ shares available in the LOB at the best bid.  To establish an analogy with second price auctions, suppose that we approximate the availability of shares at specified prices in the LOB by a density function $w(p)$ such that there is $\d V = w(p)\d p$ volume available at the price $m + p$, \ie~$p \ge 0$ is an offset from the mid price.  Then, the \textit{total} volume available at or below price $p$ is given by $W(p) = \int_0^p w(x)\d x$.  This is an unnormalized supply curve.
If we decide to buy all the available shares up to and including price $m + p$, then we will need to pay a total of $f(p) = \int_0^p x w(x)\d x$, which is exactly the expected cost incurred by bidding $p$ in a second price auction with supply curve $W$.  As well, the transaction costs associated with purchasing a volume $V$ of shares (\ie~the cost of a \textit{market order} for $V$ shares) is given by $\Lambda(V) = f \circ W^{-1}(V)$, which is exactly the second price acquisition cost function.  As a corollary to Proposition~\ref{prop:convex_acquisition_costs_2pa}, these transaction costs are convex functions of volume and hence can be incorporated into tractable portfolio construction problems.  As an example, if $w(p) = w_0 p$, then $W(p) = \frac{1}{2}w_0p^2$ and $W^{-1}(V) = \sqrt{\frac{2V}{w_0}}$ and $\Lambda(V) \propto V^{3/2}$, which reproduces the functional form of the famed square-root law~\cite[Ch. 12]{bouchaud2018trades} (but for much different reasons than the usual derivation of this law).

An important problem in finance is to construct, based on an estimate of risk and future returns, a portfolio that optimally trades off between these two aspects.  One of the first formulations of this problem was famously carried out via mean-variance optimization (quadratic programming) by Markowitz~\cite{markowitz_portfolio_selection}, and this remains an active area of research, see \eg~\cite{tuck21_portf_const_using_strat_model, lobo2007portfolio, abeille2016lqg}, and many others.

We can formulate a simple instance of this problem that takes into account order book volume costs as

\begin{equation}
  \label{eqn:markowitz_problem}
  \underset{x \in \R^M}{\text{minimize}} \quad \frac{1}{2}\lambda x^\T \Sigma x + \sum_{j = 1}^M \bigl[\Lambda_j(x_j) - \alpha_j x_j \bigr], \tag{$\mathcal{M}$}
\end{equation}
where $x \in \R^M$ is the portfolio allocation across $M$ risky assets so that $x_j \in \R$ is the number of shares of type $j$ that are purchased, $\alpha_j$ is the forecast future returns of asset $j$, and $\Sigma \succ 0$ is the covariance matrix of these returns.  This covariance matrix arises from the calculation $\Var \alpha^\T x$, which is often used as a measurement of the \textit{risk} of portfolio $x$, and is scaled by the risk-aversion parameter $\lambda > 0$.  This formulation is a convex optimization problem, the solution of which tells us how many shares of which type to buy in order to take advantage of the knowledge $\alpha_j$ of future returns, while taking into account the actual cost of acquiring those shares given the present state of the order book.  In this simple example, we include no costs associated with leverage or with borrowing shares to short.  However, these are also convex constraints, and are easily incorporated in Problem~\eqref{eqn:markowitz_problem}.

This problem admits an interesting dual problem which we may analyze by first introducing an additional variable $u \in \R^M$, and a trivial constraint $x = u$:

\begin{align*}
  \underset{x \in \R^M, u \in \R^M}{\text{minimize}} &\quad \frac{1}{2}\lambda u^\T \Sigma u + \sum_{i = 1}^M \bigl[\Lambda_j(x_j) - \alpha_j x_j \bigr]\\
  \text{subject to} &\quad x = u.
\end{align*}
Carrying out the usual calculations for the Lagrangian function, we obtain

\begin{align*}
  \L(x, u, \phi)
  &= \frac{1}{2}\lambda u^\T \Sigma u + \sum_{j = 1}^M \bigl[\Lambda_j(x_j) - (\alpha_j - \phi_j) x_j \bigr] - \sum_{j = 1}^N \phi_j u_j,
\end{align*}
and minimizing with respect to $x$, we obtain $\L(x^\star, u, \phi) = \frac{1}{2}\lambda u^\T \Sigma u - \sum_{j = 1}^M \phi_j u_j - \sum_{j = 1}^M \Lambda_j^\star(\alpha_j - \phi_j)$, which, when minimizing again over $u$ results in $u^\star = -\frac{1}{\lambda}\Sigma^{-1} \phi$, and the dual function $d(\phi) = -\frac{1}{2\lambda}\phi^\T \Sigma^{-1} \phi - \sum_{i = 1}^N \Lambda_i^\star(\alpha_i - \phi_i)$.  Incidentally, the quantity $x = -\frac{1}{\lambda}\Sigma^{-1} \alpha$ is exactly the optimal portfolio when $\Lambda_i \equiv 0$ (\cf~$u^\star$), so that the dual vector $\phi$ can be understood as a sort of \textit{cost-adjusted} forecast.

Thus, through the dual, the optimal portfolio can be obtained by solving

\begin{equation*}
\underset{\phi}{\text{minimize}} \quad \frac{1}{2\lambda} \phi^\T \Sigma^{-1} \phi + \sum_{j = 1}^M \Lambda_j^\star(\alpha_j - \phi_j).
\end{equation*}
This can be further simplified by using the Cholesky decomposition $\Sigma = LL^\T$ (and hence $\Sigma^{-1} = L^{-\T}L^{-1}$) and the change of variables $\zeta = L^{-1} \phi$ to obtain

\begin{equation*}
  \underset{\zeta}{\text{minimize}} \quad \bm{\Lambda}^\star(\alpha - L\zeta) + \frac{1}{2\lambda} ||\zeta||_2^2,
\end{equation*}
where $\bm{\Lambda}^\star(z) = \sum_{j = 1}^M \Lambda_j^\star(z_j)$.  This is now nothing but a regularized minimization problem of the Fenchel conjugates.  The advantage of this formulation is that derivatives of $\Lambda^\star$ (see Section~\ref{sec:acquisition_costs}) are in terms of the volume available in the order book (\ie~$w$ alone) without involving the inverse of $W$, which occurs in derivatives of the primal objective.

\subsection{More Related Problems}

\subsubsection{Statistical Arbitrage Mining}
The work of~\cite{zhang2015statistical} analyzes a budget constrained optimal bidding problem with multiple contracts in a context they refer to as \textit{statistical arbitrage mining} (SAM).  Rather than the contracts stipulating a hard requirement, contract $i$ pays out a \textit{random} amount $v_{ij}$ for each impression of type $j$ that is obtained.  The value is random in order to encode a probability of an impression of type $j$ resulting in a \textit{conversion} for contract $i$.  The objective is to maximize the expected profit $\sum_{j = 1}^M \lambda_j W_j(x_j)\sum_{i = 1}^N \gamma_{ij} \E v_{ij} - \sum_{j = 1}^M \lambda_j f_j(x_j)$ over the bids $x \in \R^M$ and probability vectors $\gamma_{ij} \ge 0$ with $\sum_{i = 1}^N \gamma_{ij} = 1$.  The variable $\gamma_{ij}$ indicates the proportion of items of type $j$ to be allocated towards contract type $i$, in exact analogy to the Problem~\eqref{eqn:main_problem_Wf}.  The SAM problem is subject to a budget constraint $\sum_{j = 1}^M \lambda_j f_j(x_j) \le B$ and has an additional variance constraint $\Var \Bigl(\sum_{j = 1}^M \lambda_j W_j(x_j) \sum_{i = 1}^N \gamma_{ij} v_{ij} \Bigr) \le h$.  The variance is taken over the randomness in $v_{ij}$.  This variance constraint can be thought of as a \textit{risk penalty}, in analogy to the Markowitz portfolio problem.  It is shown by~\cite{zhang2015statistical} that the bids $x_j$ and allocation array $\gamma_{ij}$ are tuned to take advantage of discrepancies between payouts and $v_{ij}$ and true item valuations.  This problem can also be recast as a convex program.

\subsubsection{The Dark Pool Liquidation Problem}
Another interesting example from finance, the \textit{dark pool problem}~\cite{ganchev2010censored}, is the problem of how to allocate shares across multiple ``dark pool'' (DP) exchanges in order to maximize the number of shares that are sold.  A DP is a type of financial exchange providing an alternative to the limit order book.  In a DP, prices and volumes are not quoted, rather, one simply announces their willingness to transact a certain volume of securities and agree to a transaction, if there is an available counter party, at the mid price quoted on some exogenous LOB exchange.  The purpose of these markets is to enable institutions to liquidate large blocks of shares without revealing this large order to the rest of the market, and hence adversely impacting the price.  It can be seen that the objective function of the problem considered by~\cite{ganchev2010censored} involves terms equivalent to $\E{(x - \xi)}_+$, where $\xi$ is a random variable and $x$ is a decision variable.  A connection with auction theory arises through the fact that $\E{(x - \xi)}_+ = f_{1st}(x) - f_{2nd}(x)$ and as well that $\E{(x - \xi)}_+ = \Lambda^\star_{2nd}(x)$.

\section{Conclusion}
\label{sec:conclusion}
This paper has focused on the implications of the convexity inducing transformation $s = \lambda W(x)$ for optimum bidding problems.  The essence of this transformation is to change problems of optimizing the bid $x$ into a problem of optimizing the probability of winning an item, \ie~it is a \textit{quantile transformation}.  The perspective offered by this transformation of variables plays a key role in reformulating the RTB contract management problem~\cite{marbach_bidding_2020, kinnearbidding2_2020} into the tractable convex program~\eqref{eqn:main_problem}.  It was seen that this convex transformation applies unconditionally in second price auctions, and Section~\ref{sec:first_price_case} shows that the weak condition of $2$-concavity of $W$ (log-concavity also being sufficient) is enough to ensure convexity in the first price case as well.

The implications of duality for Problem~\eqref{eqn:main_problem} are incredibly rich.  This has been summarized in Proposition~\ref{prop:duality_consequences} where we saw that optimal bids $x_j$ can be fully characterized by solutions $(\rho, \mu)$ of the dual Problem~\eqref{eqn:dual_problem}.  This can be done either directly through the dual multipliers $\mu_j$ associated to item type constraints $j \in [M]$, or indirectly through the \textit{pseudo-bids} $\rho_i$ associated to contract constraints $i \in [N]$.  As well, the dual variables of the dual program (\ie~the double dual variables) reproduce the optimum allocation array $R$ for the primal problem.  We have also seen from the dual program that the slack variables $\theta_{ij} = \mu_j - v_{ij}\rho_i$ induce sparsity in the allocation array $R$ in the sense that $\theta_{ij} > 0 \implies R_{ij} = 0$.  Since Problem~\eqref{eqn:main_problem} is convex, there are scalable and efficient algorithms available for its practical solution and some illustrative computational examples are given in Section~\ref{sec:computational_examples}.

In Section~\ref{sec:related_problems} we applied the convexifying transformation $s = \lambda W(x)$ to existing problems in the literature, particularly budget constrained optimum bidding problems considered in~\cite{zhang2014optimal, zhang2015statistical, balseiro2015repeated}, demonstrating that these problems can similarly be re-cast as convex programs.  We also drew, in Section~\ref{sec:lob_problem}, an equivalence between the cost of winning items with a target probability in second price auctions (\ie~the acquisition cost function) and the cost of submitting market orders in limit order book based financial markets.  The convexity of this function is used to incorporate a certain type of transaction costs into a Markowitz portfolio problem.  The dual of this problem is equivalent to a Tikhonov regularized minimization problem of the conjugate function $\Lambda^\star$, the derivatives of which are more easily available (\cf~Remark~\ref{rem:derivatives}).  Connections to statistical arbitrage mining~\cite{zhang2015statistical} and the dark pool problem~\cite{ganchev2010censored} are also remarked upon.

Given the examples discussed in Section~\ref{sec:related_problems}, it appears that there are many other important applications arising in finance and auction theory that could be studied or re-cast through the lens of convex analysis.

\clearpage
\printbibliography
\clearpage

\appendix
\section{Proofs}
\label{sec:appendix}
\begin{theoremEnd}[normal]{lemma}
  \label{lem:ubp}
  In a first or second price auction, suppose that the acquisition cost curves $\Lambda_j$ are convex.  Then, if a solution $x_{ij}, \gamma_{ij}$ to Problem~\eqref{eqn:main_problem_Wf} exists, there is also a solution with the property that $\forall i \in [N]:\ x_{ij} = x_j$ and such that $\sum_{i \in \B_j} \gamma_{ij}(t) \in \{0, 1\}$.
\end{theoremEnd}
\begin{proofEnd}
  Suppose $(x, \gamma)$ is a solution of Problem~\eqref{eqn:main_problem}, and with total cost $J$. Let $(\tilde{x}, \tilde{\gamma})$ be another candidate solution with total cost $\tilde{J}$ defined by
  \begin{align*}
    \tilde{x}_j
    &\defeq W_j^{-1}\Bigl(\sum_{i = 1}^N\gamma_{ij}W_j(x_{ij})\Bigr),\\
    \tilde{\gamma}_{ij}
    &\defeq \frac{\gamma_{ij} W_j(x_{ij})}{\sum_{u = 1}^N \gamma_{uj} W_j(x_{uj})},
  \end{align*}
  where $0 / 0 \defeq 0$ in the definition of $\tilde{\gamma}$.  We proceed to show that $(\tilde{x}, \tilde{\gamma})$ is also a solution and we note that the definition of $\tilde{\gamma}$ satisfies $\sum_{i = 1}^N\tilde{\gamma}_{ij} \in \{0, 1\}$. 

  The pair $\tilde{x}, \tilde{\gamma}$ is feasible by construction.  Indeed, $\tilde{\gamma}_{ij} \ge 0$ and $\sum_{i = 1}^N \tilde{\gamma}_{ij} \le 1$ by definition. Moreover, we have

  \begin{align*}
    \sum_{j = 1}^M \tilde{\gamma}_{ij} \lambda_j v_{ij} W_j(\tilde{x}_j)
    &= \sum_{j = 1}^M \Bigl[\frac{\gamma_{ij} \lambda_j v_{ij} W_j(x_{ij}) \sum_{v = 1}^N\gamma_{vj}W_j(x_{vj})}{\sum_{u = 1}^N\gamma_{uj}W_j(x_{uj})} \Bigr]\\
    &= \sum_{j = 1}^M \gamma_{ij} \lambda_j v_{ij} W_j(x_{ij})\\
    &= C_i,
  \end{align*}
  where the last equality follows since $x, \gamma$ is assumed to be a solution.
  
  Now, we see that the cost of $(\tilde{x}, \tilde{\gamma})$ satisfies $\tilde{J} = J$ since $J$ is the minimal cost and
  \begin{align*}
    \tilde{J}
    &\defeq \sum_{i = 1}^N \sum_{j = 1}^M \tilde{\gamma}_{ij} f_j(\tilde{x}_j) \\
    &\overset{(a)}{=} \sum_{i = 1}^N \sum_{j = 1}^M \tilde{\gamma}_{ij} \Lambda_j\Bigl(\sum_{u \in \B_j}\gamma_{uj}W_j(x_{uj}), t\Bigr)\\
    &\overset{(b)}{\le} \sum_{i = 1}^N \sum_{j = 1}^M \tilde{\gamma}_{ij} \sum_{u = 1}^N\gamma_{uj} \Lambda_j(W_j(x_{uj}))\\
    &\overset{(c)}{=} \sum_{j = 1}^M\sum_{u = 1}^N \gamma_{uj} f_j(x_{uj}) \sum_{i = 1}^N \tilde{\gamma}_{ij} = J\\
  \end{align*}
  where $(a)$ is just the definition of $\Lambda_j = f_j \circ W_j^{-1}$ (see Section~\ref{sec:cost_functions}), $(b)$ follows by the convexity of $\Lambda_j$ and that $\Lambda_j(0) = 0$ (since $\gamma_{ij}$ need not necessarily sum to $1$), and $(c)$ follows again by $\Lambda_j = f_j \circ W_j^{-1}$ and then by interchanging the order of summation.

  Since $\tilde{x}, \tilde{\gamma}$ is feasible and has optimal cost, it is a solution.  
\end{proofEnd}

\begin{theoremEnd}[normal]{proposition}[Hierarchy of $\alpha$-concavity]%
  \label{prop:alpha_log_concavity}
  For $0 \le \alpha < \beta$, if $W$ is $\alpha$-concave, then it is
  also $\beta$-concave.
\end{theoremEnd}
\begin{proofEnd}
  First we check that $\ell_\beta \circ \ell_\alpha^{-1}(x)$ is both
  monotone increasing and concave.  This follows if the first
  derivative

  \begin{equation*}
    \frac{\d}{\d x}\ell_\beta \circ \ell_\alpha^{-1}(x)
    = \frac{\ell_\beta' \circ \ell_\alpha^{-1}(x)}{\ell_\beta' \circ \ell_\alpha^{-1}(x)},
  \end{equation*}
  is positive and monotone non-increasing; which, since
  $\ell_\alpha^{-1}$ is itself monotone non-decreasing, follows if

  \begin{equation*}
    \frac{\ell_\beta'(x)}{\ell_\alpha'(x)}
  \end{equation*}
  is both positive and monotone decreasing.  Since
  $\ell_\alpha'(x) = x^{-\alpha}$ this function is
  $x^{\alpha - \beta}$, which is positive on the domain $x > 0$ and
  decreasing if $\alpha < \beta$.

  Now, we check concavity of $\ell_\beta \circ W$ directly from the
  definition, for $t \in (0, 1)$:

  \begin{align*}
    \ell_\beta \circ W(tx + (1 - t) y)
    &= \ell_\beta \circ \ell_\alpha^{-1} \circ \ell_\alpha \circ W(tx + (1 - t)y)\\
    &\overset{(a)}{\ge} \ell_\beta \circ \ell_\alpha^{-1}\bigl(t\ell_\alpha \circ W(x) + (1 - t)\ell_\alpha \circ W(y) \bigr)\\
    &\overset{(b)}{\ge} t\ell_\beta \circ \ell_\alpha \circ \ell_\alpha^{-1} \circ W(x) + (1 - t)\ell_\beta \circ \ell_\alpha^{-1} \circ \ell_\alpha \circ W(y)\\
    &= t \ell_\beta \circ W(x) + (1 - t) \ell_\beta \circ W(y),
  \end{align*}
  where $(a)$ follows by the assumed concavity of
  $\ell_\alpha \circ W$ and the monotonicity of
  $\ell_\beta \circ \ell_\alpha^{-1}$ while $(b)$ from the concavity
  of $\ell_\beta \circ \ell_\alpha^{-1}$.  
\end{proofEnd}

% NOTE: There seems to be a bug with proof-at-the-end that depends
% on how many proofs there are?  Putting proofs in multiple different
% categories (main, main2,... aux, aux2, ...) seems to fix the problem.
\end{document}